\documentclass[10pt,letterpaper,twocolumn]{article}

\DeclareUnicodeCharacter{03B1}{$\alpha$}
\DeclareUnicodeCharacter{03B2}{$\beta$}
\DeclareUnicodeCharacter{0391}{$\mathrm{A}$}

\usepackage[margin=0.75in]{geometry}
\usepackage[T1]{fontenc}
\usepackage[utf8]{inputenc} 
\usepackage{newtxtext}
\usepackage{newtxmath}

\usepackage{titlesec}

\titleformat{\section}
  {\centering\bfseries\scshape}
  {\Roman{section}.}
  {0.6em}
  {\MakeUppercase}

\titleformat{\section}
  {\centering\scshape}
  {\Roman{section}.}
  {0.6em}
  {\MakeUppercase}

\titleformat{\subsection}
  {}
  {\Alph{subsection}.}
  {0.6em}
  {\itshape}

\titlespacing*{\section}{0pt}{1.2ex plus 0.5ex minus 0.2ex}{0.8ex plus 0.2ex}
\titlespacing*{\subsection}{0pt}{1.0ex plus 0.4ex minus 0.2ex}{0.6ex plus 0.2ex}

\usepackage{amsmath}
\usepackage{bm}

\usepackage{graphicx}
\usepackage{booktabs}
\usepackage{float}
\usepackage{array}
\usepackage{makecell}

\usepackage{url}
\usepackage{textcomp} 
\usepackage{soul}
\usepackage{color}
\usepackage{verbatim}
\usepackage[none]{hyphenat}

\usepackage[
  backend=biber,
  style=ieee,
  doi=false,
  isbn=false,
  url=false,
  eprint=false,
  date=year
]{biblatex}
\AtEveryBibitem{%
  \clearfield{urldate}%
  \clearfield{note}%
  \clearfield{month}%
  \clearfield{day}%
  \clearlist{language}%
}

\newcommand{\rework}[1]{}

\makeatletter

\newcommand{\Rom}[1]{\expandafter\@slowromancap\romannumeral #1@}
\makeatother

\title{\bfseries\LARGE
Interfacial Strain and Structural Defects Govern the Performance of Tantalum Superconducting Waveguide Resonators
}

\author{%
\parbox{\textwidth}{%
\centering
M. Singer$^{1}$, H. Gupta$^{1}$, B. Schoof$^{1}$, E. Willinger$^{2}$, A. Orekhov$^{2}$, M. Tornow$^{1*}$\\[0.8em]
\raggedright
$^{1}$School of Computation, Information, and Technology, Technical University of Munich, Germany\\[0.4em]
$^{2}$Electron Microscopy Facility, Technical University of Munich, Germany\\[0.4em]
$^*$ Corresponding author: tornow@tum.de
}%
}

\date{}

\begin{document}

\twocolumn[{
\maketitle

\begin{abstract}

Tantalum (Ta) is a promising material for reaching long coherence times in superconducting qubits. A detailed understanding of the underlying structure–property relationship remains elusive though. In the present study, we sputter-deposited 200 nm thick Ta films on high-resistivity silicon (100) substrates at temperatures ranging from T = 20\textdegree C to 600\textdegree C, as well as on different seed layers (Nb, TiN and TaN). Alpha-Ta thin films were readily obtained at temperatures above 500\textdegree C and on all seed layers. The films were characterized in terms of surface morphology, residual-resistance ratio, crystal phase composition and superconducting transition temperature, as well as RF-performance using coplanar waveguide resonators. Internal quality factors of up to 1.5 million were measured at 100 mK in the single-photon regime. Despite similar bulk material properties, alpha-Ta films on different seed layers exhibit markedly different RF performance, which we attribute to dissimilar strain and structural defects at the substrate–metal interfaces. Williamson-Hall analysis of XRD data reveals a clear correlation between decreasing microstrain and increasing quality factor. Cross-sectional HR-TEM further supports this interpretation by directly resolving interfacial disorder. Our results highlight the critical role of interface engineering in optimizing superconducting thin films for low-loss quantum computing circuitry.

\end{abstract}

\vspace{2em}
}]

\section{INTRODUCTION}

 Superconducting thin-film circuit technology is one of today's prevailing platforms to realize large-number qubit systems for future quantum computing applications \cite{KRANTZ2019QuantumEngineerGuideSuperconductingQubits}. While largely scalable by using well-established standard semiconductor processes for fabrication, much of recent years' efforts have focused on the materials aspects as it has become increasingly obvious how material related losses may critically limit the maximal achievable coherence times  \cite{huang_superconducting_2020}.

In this context, not only the actual superconductor thin-film materials used for the circuitry elements (base plate, resonators, capacitors, flux lines, Josephson junctions) but significantly also, the various involved surfaces and interfaces, need to be taken into account if optimized performance is to be achieved \cite{ALTOE2022LocalizationMitigationLossNiobiumSuperconductingCircuits, MURRAY2021Materialmatterssuperconductingqubits}.
The dominating material standard of leading academic labs and industrial fabrication is based on aluminum and niobium, while in recent years tantalum (Ta) has strongly emerged as an alternative material of high promise, due to high reported resonator quality factors and qubit coherence times
\cite{PLACE2021Newmaterialplatformsuperconductingtransmonqubitscoherencetimesexceedingmilliseconds, tennantPRXQuantum}. 
While several factors would contribute to this apparent material advantage of Ta, it is assumed that in particular the about three nanometers thick native Ta oxide is more stable and stoichiometric, and therefore hosts less Two Level Systems (TLSs) compared to, e.g., the native oxide of niobium \cite{OH2024StructureFormationMechanismsTantalumNiobiumOxidesSuperconductingQuantumCircuits, wang_why_2025}. At the cryogenic temperatures of qubit operation, these TLSs are often identified as the limiting factor as they couple to the microwave field of qubit operation, resulting in dielectric losses and consequently to reduced coherence times. The origin of these TLSs is not fully understood, yet, but it is anticipated that they originate from tunneling atoms, dangling bonds, hydrogen defects or collective motions of small atomic groups - all predominantly found in the surfaces and interfaces of superconducting circuits. \cite{MULLER2019understandingtwolevelsystemsamorphoussolidsInsightsquantumcircuits, PHILLIPS1987Twolevelstatesglasses, WOODS2019DeterminingInterfaceDielectricLossesSuperconductingCoplanarWaveguideResonators}. 

The interplay of TLS losses with existing strain in superconducting thin films, and as a consequence, with the RF performance of resonators fabricated from these, has been described before \cite{GRABOVSKIJ2012StrainTuningIndividualAtomicTunnelingSystemsDetectedSuperconductingQubit}, but no systematic study involving a detailed structure-property correlation, in particular for Ta thin films, has been reported, to the best of our knowledge.

Tantalum has two crystalline phases: the stable bcc-phase, which is also called alpha-Ta, and the meta-stable tetragonal phase beta-Ta \cite{KLAVER2002ThinTafilmsgrowthstabilitydiffusionstudiedmoleculardynamicssimulations}. To prepare Ta thin films, mostly sputter deposition techniques have been employed, and the impact of various sputtering parameters on the resulting Ta phase has been studied before \cite{CATANIA1993Phaseformationmicrostructurechangestantalumthinfilmsinducedbiassputtering,ZHOU2009Effectsdepositionparameterstantalumfilmsdepositeddirectcurrentmagnetronsputtering,SCHAUER1972SPUTTERED13TANTALUMTANTALUMFILMS, INO1997IonenergyionfluxionspecieseffectscrystallographicelectricalpropertiessputterdepositedTathinfilms, HOOGEVEEN1996TexturephasetransformationsputterdepositedmetastableTafilmsTaCumultilayers}. It was reported that the alpha-phase can form under a certain sputter chamber pressure \cite{NAVID2012NanostructuredalphabetatantalumformationRelationshipplasmaparametersmicrostructure} and that post-deposition annealing can transition beta- to alpha-tantalum \cite{CLEVENGER1992relationshipdepositionconditionsbetaalphaphasetransformationstressrelaxationtantalumthinfilms,KNEPPER2006Effectoxygenthermomechanicalbehaviortantalumthinfilmsphasetransformation,FEINSTEIN1974AnnealingphasestabilitytantalumfilmssputteredArO2, LIU2001AnnealingeffectstantalumfilmsSiSiO2Sisubstratesvariousvacuums, LEE2004Texturestructurephasetransformationsputterbetatantalumcoating}. In general however, there are two well-established methods to obtain alpha-tantalum. The first one is to deposit the tantalum thin films at elevated temperatures above 400-500°C \cite{ZHANG2006Hardnessenhancementnanocrystallinetantalumthinfilms}. Tantalum films sputtered at high temperatures and therefore predominantly consisting of the alpha-phase were used to fabricate qubits with lifetimes of up to 0.5 ms on sapphire substrates \cite{WANG2022practicalquantumcomputerstransmonqubitlifetimeapproachingmilliseconds}. Superconducting coplanar waveguide (CPW) resonators from high-temperature sputter-deposited Ta on silicon  \cite{Lozano_2024} and sapphire \cite{CROWLEY2023DisentanglingLossesTantalumSuperconductingCircuits} were investigated as well, with quality factors reaching 15 million in the single-photon regime \cite{BLAND2025Millisecondlifetimescoherencetimes2Dtransmonqubits}. The second method for obtaining alpha-Ta is to use a few nanometers thick seed layer below Ta, which nucleates the formation of the alpha-Ta phase \cite{SINGER2024TantalumThinFilmsSputteredSiliconDifferentSeedLayersMaterialCharacterizationCoplanarWaveguideResonatorPerformance}.  Suitable seed layers, on which alpha-phase formation has been demonstrated, are tantalum-nitride, niobium and titanium. Resonators built from tantalum thin films that were deposited on a niobium seed layer on silicon  \cite{urade_microwave_2024, MARCAUD2025Lowlosssuperconductingresonatorsfabricatedtantalumfilmsgrownroomtemperature}  have already been demonstrated and showed promising results (internal quality factors in the range of $10^6$).

Thus, previous work has revealed that Ta is a highly appealing alternative to Nb for fabricating resonators and further qubit components. Still, there is only limited information from systematic studies available, including a comparison of alpha-Ta films that were obtained by high-temperature deposition vs. those formed by nucleation on different seed layers  \cite{alegria_two-level_2023-1}. 
In particular, a detailed advancement of understanding of which material composition, process or fabrication steps would exactly cause the apparent increase in TLS-originated dielectric losses at the interfaces, and why they do so, seems to be largely missing.

\begin{figure}[h!t]
  \centering
  \includegraphics[width=\columnwidth]{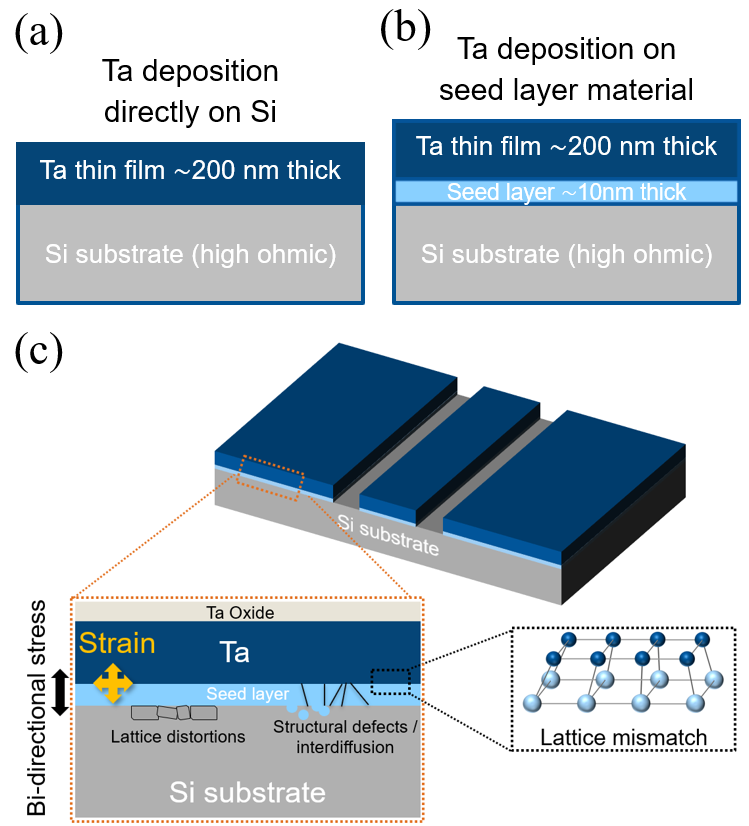}
  \caption{Sketch of the material stacks in the case of Ta thin-films directly deposited on Si substrates (a) and on a seed layer from TiN, TaN or Nb (b). (c) Schematics of part of a CPW resonator (top panel) and a cross-section through its material stack, indicating various possible loss-inducing imperfections such as structural defects, material interdiffusion, lattice mismatch and distortions, all potentially contributing to bi-directional stress across the layer stack and resulting in built-in strain at the substrate-to-metal interface (bottom panel).}
  \label{Cartoon}
\end{figure}

Therefore, in the present work, we set out to undertake a detailed and systematic study, by depositing Ta thin-films at different temperatures ranging from 20°C to 600°C and specifically, by using Nb, tantalum-nitride (TaN), and titanium-nitride (TiN) thin-films as seed layers for Ta deposition, all on silicon (100) substrates, as shown schematically in Fig.~\ref{Cartoon}(a)+(b). In general, the various material combinations and sputter conditions may result in dissimilar lattice mismatches and distortions, material defects and interdiffusion, and different strain situations at the interfaces, as illustrated in Fig.~\ref{Cartoon}(c). We characterized the surface roughness of these films with Atomic Force Microscopy (AFM) and their electrical thin-film parameters in terms of resistivity, residual resistance-ratio (RRR) and critical temperature. The crystallography of the films was investigated with grazing incidence X-ray diffraction (GI-XRD) and the phase composition (alpha vs. beta) was extracted from these data. The films were then used to fabricate CPW resonators in the 5-9 GHz regime, and the internal quality factors were measured at different input powers at a temperature of 100 mK. Our findings both clearly discriminate the process conditions under which the favorable alpha-phase of Ta can be obtained. Strikingly however, it turns out that preparing alpha-Ta alone does not guarantee resonator high-performance, yet. Rather, our analysis of the XRD data together with cross-sectional TEM imaging directly reveal that micro-strain and lattice distortions present in the interfacial layers, as apparent when TaN is the seed layer, strongly correlate with degraded RF performance - \textit{even if alpha-Ta films were realized}. Conversely, using Nb as seed layer to grow alpha-Ta guarantees a minimum of lattice distortions in the underlying Si substrate and lowest built-in microstrain in the Ta-to-substrate interface region - enhancing the measured internal quality factors to their maximum values of about 1.5 million in the single-photon regime, at 100 mK.

We expect our insights to make high-quality Ta thin-film deposition for superconducting qubit devices readily available such that this material may further establish itself as strong future alternative to currently more common superconductors.   

\section{Results} 


\subsection{Surface and crystallographic characterization of tantalum thin films}

The measured roughness of the various thin films varies with sputter conditions. The tantalum sputtered directly on silicon at room temperature (20°C) has an RMS roughness of 0.5~$\pm$ 0.1~nm, while the thin film sputtered on the 600°C heated substrate shows a higher roughness of 1.5~$\pm$ 0.1~nm and more elongated structures (for the AFM images, see Supplementary Fig.~S1). When Nb is used as a seed layer, the roughness has a lower value of about 0.7 $\pm$ 0.1 ~nm. Tantalum sputtered on titanium nitride and tantalum nitride both show similar RMS roughness values of 0.8 $\pm$ 0.1 ~nm. The surfaces of the Ta on Nb, TiN and TaN also appear to be of a similar morphology.

\begin{figure}
  \includegraphics[width=0.5\textwidth]{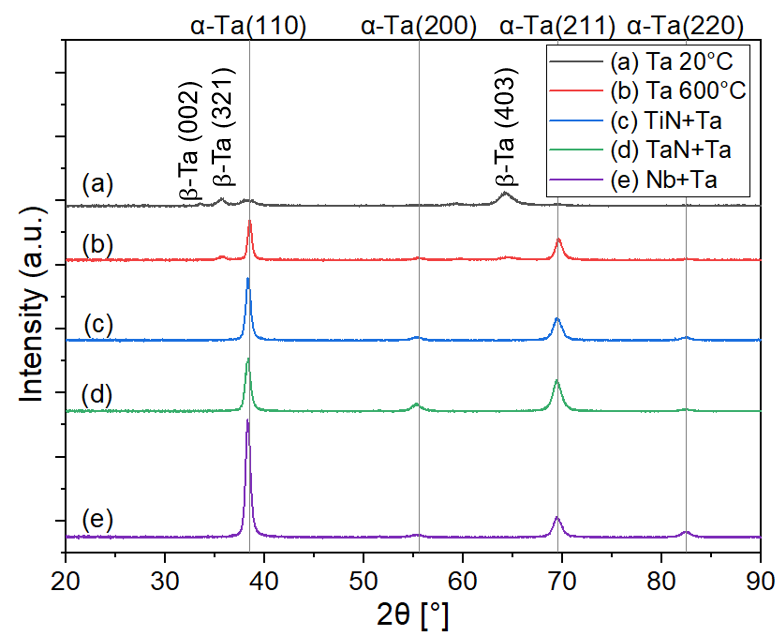}
  \caption{GI-XRD spectra taken at 0.5° incidence beam angle for tantalum thin films sputtered at different temperatures directly on the silicon substrate or sputtered at 20°C on ca. 10~nm thick seedlayers from either TiN, TaN or Nb.  Vertical lines indicate expected peak positions corresponding to selected alpha-tantalum crystal planes \cite{HALLMANN2013Effectsputteringparameterssubstratecompositionstructuretantalumthinfilms, HOOGEVEEN1996TexturephasetransformationsputterdepositedmetastableTafilmsTaCumultilayers, KNEPPER2006Effectoxygenthermomechanicalbehaviortantalumthinfilmsphasetransformation, MYERS2013phasetransitiontantalumcoatingsdepositedmodulatedpulsedpowermagnetronsputtering}.}
  \label{XRD}
\end{figure}

The GI-XRD spectra for the various Ta films under study are shown in Fig.~\ref{XRD}. The black curve at the top (a) represents Ta sputtered on a silicon (100) substrate at room temperature (20°C). In this spectrum, a peak is visible at the position of alpha-Ta (110), which however is not very pronounced. Besides this peak, there is also a very small peak around 33.5° which can be attributed to beta-Ta (002). There is another small peak at 35.6° which corresponds to beta-Ta (321). Finally, one more pronounced peak at 64.3° can be assigned to the position of beta-Ta (403) \cite{HALLMANN2013Effectsputteringparameterssubstratecompositionstructuretantalumthinfilms, HOOGEVEEN1996TexturephasetransformationsputterdepositedmetastableTafilmsTaCumultilayers, KNEPPER2006Effectoxygenthermomechanicalbehaviortantalumthinfilmsphasetransformation, MYERS2013phasetransitiontantalumcoatingsdepositedmodulatedpulsedpowermagnetronsputtering}. The spectrum indicates that sputtering Ta on Si (100) without heating of the substrate results in a mixed-phase Ta film with beta-Ta dominating. The red curve (b) below shows the diffraction spectrum for Ta sputtered on the 600°C heated Si substrate. In contrast to the unheated sputtered Ta, the diffraction pattern shows only a reduced beta‑Ta (321) peak, with the beta‑Ta (002) and (403) peaks almost completely suppressed. Instead, alpha-Ta peaks corresponding to (110) and (211) are clearly visible. This indicates that, as reported already before, heating influences the phase formation in Ta thin films, with alpha-Ta being favored at high deposition temperatures. This is further supported by the diffraction spectra of Ta sputtered at 100°C, 200°C, 300°C, 400°C and 500°C that are shown in the Supplementary Information, Fig.~S2. The three spectra below belong to the Ta thin-films on seed layers with blue (c) for Ta on TiN, green (d) for Ta on TaN and purple (e) for Ta on Nb. In all of these spectra peaks only appear at the expected alpha-Ta positions. The peak for alpha-Ta (110) is the most prominent one in all of the films, while it is most pronounced for Ta on Nb. The second sharp peak in the three spectra is from the alpha-Ta (211) plane. Alpha-Ta (200) and (220) peaks are also visible in all three spectra for Ta on seed layer, but they are markedly weaker and much less distinct than the (110) and (211) peaks. In summary, for all Ta on seed‑layer films studied here, the chosen seed materials consistently nucleate the growth of the alpha‑Ta phase.


\subsection{Electrical characterization of tantalum thin films}

\begin{figure*}[h!t]
  \includegraphics[width=\textwidth]{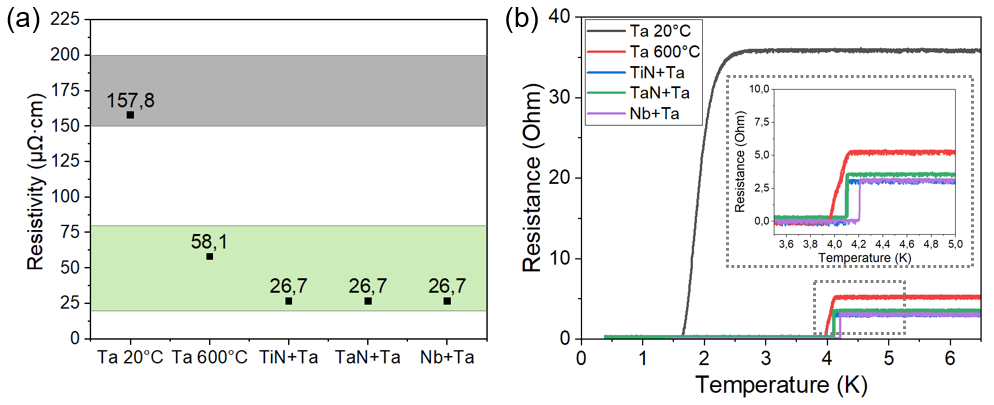}
  \caption{Electrical characterization of sputtered tantalum thin films. (a) Room temperature resistivity values obtained from Hall-bar resistance measurements. The green band (20-80~$\mu\Omega\cdot$cm) indicates the literature value range for alpha-Ta and the grey band (150-200~$\mu\Omega\cdot$cm) the one for beta-Ta \cite{MYERS2013phasetransitiontantalumcoatingsdepositedmodulatedpulsedpowermagnetronsputtering,HALLMANN2013Effectsputteringparameterssubstratecompositionstructuretantalumthinfilms,NASAKINA2016Formationalphabetatantalumvariationmagnetronsputteringconditions,READ1965NEWSTRUCTURETANTALUMTHINFILMS}. (b) Resistance as function of temperature between 400~mK and 7~K, measured with an applied constant current of 100~$\mu$A. The Ta film comprising predominantly the beta-phase shows a lower critical temperature in accord with literature \cite{urade_microwave_2024, RAIRDEN1964Criticaltemperatureniobiumtantalumfilms}. The superconducting transition for predominantly alpha-Ta containing films is mostly sharp and well-defined while the beta-Ta thin film shows a broader transition.}
  \label{DC}
\end{figure*}

The resistivity at room temperature of all prepared Ta thin-films was determined using a Hall-bar structure (see Supplementary Information, Fig.~S3). The results of these measurements are summarized in Fig.~\ref{DC}(a), where the green band indicates the range of typical (reported) values for alpha-Ta and the gray band for beta-Ta \cite{HALLMANN2013Effectsputteringparameterssubstratecompositionstructuretantalumthinfilms,NASAKINA2016Formationalphabetatantalumvariationmagnetronsputteringconditions,READ1965NEWSTRUCTURETANTALUMTHINFILMS}. All measured resistivity values fall into two major ranges: Ta on Nb, TiN and TaN have all lowest resistivities around 26.7~$\pm$~0.1~$\mu\Omega\cdot$cm, which is in the lower range for alpha-Ta values reported in literature and agrees well with the results from the XRD measurements. The resistivity value for Ta sputtered on the 600°C heated Si is somewhat higher (58.1~$\pm$~2.0~$\mu\Omega\cdot$cm) but still in the indicated alpha-Ta range. This also matches the diffraction results, as these indicated that in this film there is mostly alpha-Ta with little remaining beta-Ta content, which might explain the somewhat higher resistivity value. In contrast to all those films, the Ta sputtered on the unheated 20°C substrate has a significantly higher resistivity value of 157.8~$\pm$~1.7~$\mu\Omega\cdot$cm, which clearly lies in the reported beta-Ta region. This is consistent with the XRD data, which indicate a mixed‑phase film with beta‑Ta as the dominant component. 

Fig.~\ref{DC}(b) displays the temperature dependence of the various thin-films' resistances, in the region around their expected transition to the superconducting state. Firstly, it can be seen that all of the investigated tantalum films turn superconducting. The Ta sputtered at 20°C on the unheated Si substrate, shows the highest residual resistance before the transition to the superconducting state. The transition to superconductivity does not occur abruptly at one single temperature but rather a smooth transition between about 2~K and 1.8~K is visible. The remaining four Ta thin films turn superconducting at higher temperatures in the range 4.0-4.2~K, which agrees with the literature value for bulk alpha-Ta \cite{IDCZAK2020TypeIIsuperconductivitycoldrolledtantalum}. Among these films, the superconducting transition of the Ta sputtered at 600°C on Si takes place over a finite temperature range, which leads to the apparent slope of the curve. Compared to the transition of the Ta sputtered at 20°C, the one of Ta 600°C exhibit a noticeably steeper transition. By comparison, the Ta thin films on TiN and TaN seed layers have a sharp transition at 4.1~K, while the Ta on the Nb seed layers has a sharp transition at an even higher temperature of 4.2~K. All $T_C$ values are in accordance with previous reports \cite{urade_microwave_2024, RAIRDEN1964Criticaltemperatureniobiumtantalumfilms}, indicating that Ta sputtering on Si at 20°C  results in a mixed, beta-rich phase, Ta sputtering on Si at 600°C results in already mostly alpha-Ta, and Ta sputtering on top of Nb, TiN or TaN in practically pure alpha-Ta thin films. 

From the room temperature resistance values and the resistance just above the superconducting transition, the RRRs of all thin films were calculated. The tantalum thin film with the high beta-Ta content (Ta 20°C) does not show any measurable variation of resistance from room to low temperature, which means that RRR\textsubscript{Ta20°C} is $\thicksim1$. In contrast, the resistance of the films with high or pure contents of alpha-Ta is decreasing from room temperature towards the superconducting transition. The tantalum thin film sputtered at 600°C has a room temperature resistance that is almost three times as high as the value before the transition, resulting in an RRR\textsubscript{Ta600°C} of 2.92. The tantalum sputtered on the Nb seed layer has an RRR\textsubscript{Nb+Ta}~=~2.20, on the TiN seed layer RRR\textsubscript{TiN+Ta}~=~2.19 and on the TaN seed layer RRR\textsubscript{TaN+Ta}~=~2.11. In general, the RRR value is determined by the charge carrier scattering at defects and impurities present in the film \cite{FELLMUTH2021Residualresistanceratioindicatorinfluenceimpuritiesfixedpointtemperatures}. Whenever such scattering dominates over (temperature dependent) scattering by phonons, RRR approaches 1. In contrast, less impurity scattering leads to smaller low temperature residual resistances and therefore usually to higher RRR values. From our results we conclude that for predominantly beta-phase Ta strong contributions from defect or impurity scattering dominates, in contrast to the Ta (mostly) alpha-phase films.


\subsection{Resonator measurements of tantalum thin films}

\begin{figure*}[!t]
  \includegraphics[width=\textwidth]{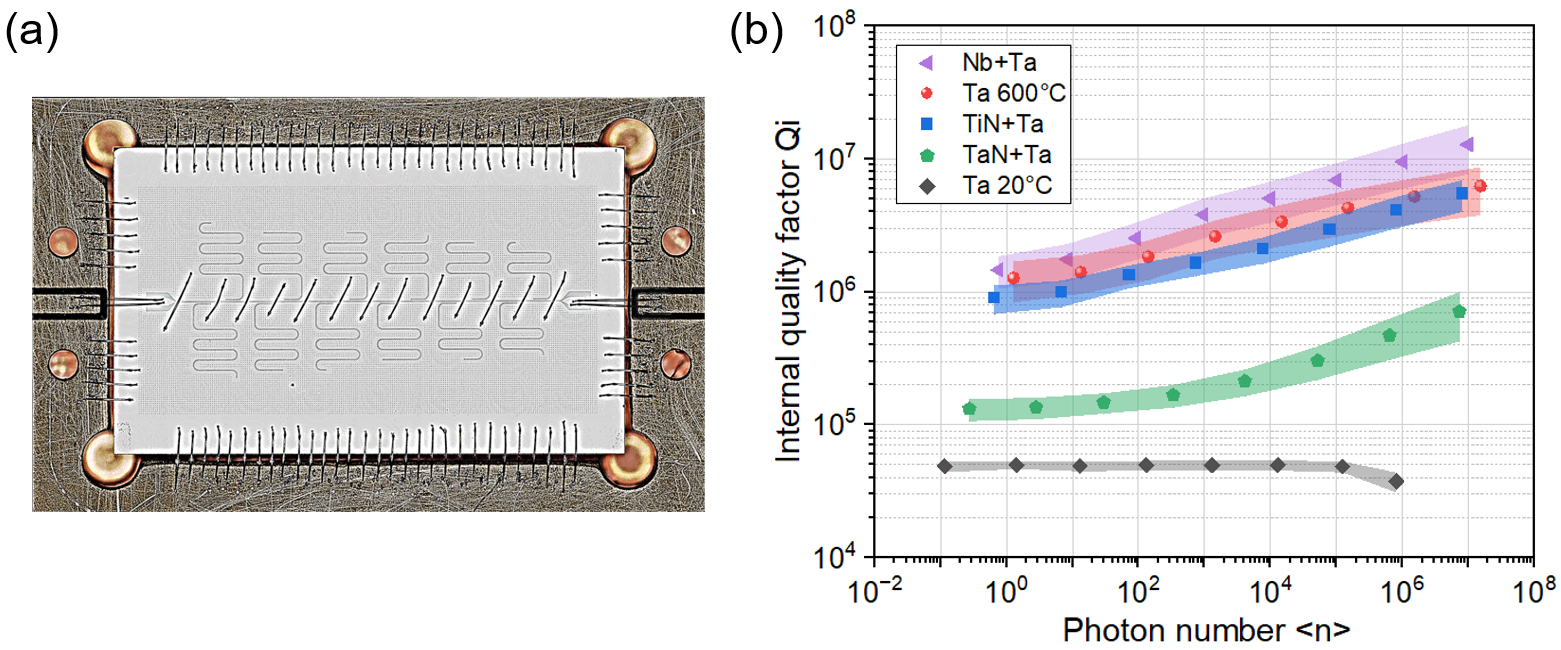}
  \caption{CPW resonator measurements of sputtered tantalum thin films. (a) a prepared and wire-bonded tantalum chip that was sputtered at 600°C. (b) comparison of internal quality factors $Q_i$ of the different tantalum thin films. The Ta thin film sputtered at 20°C on Si has the lowest $Q_i$ value while the Ta on Nb seed layer has the highest value reaching up to 1.5$\cdot10^6$ in the single-photon regime. The displayed dotted values are the mean values of all twelve resonators on the chip. The error band represents the $95\%$-confidence interval.}
  \label{Ta_RF}
\end{figure*}

The internal quality factors $Q_i$ of all different tantalum thin films under study were measured as function of RF input power and at \textit{T}=100 mK. A prepared resonator sample is shown in Fig.~\ref{Ta_RF}(a). 

As can be seen in the $Q_i$ versus photon number plot (Fig. \ref{Ta_RF}(b)), resonators fabricated from Ta deposited on Si at 20°C feature the lowest and almost power-independent internal quality factor ($Q_{i,Ta20^{\circ}C}\approx5\cdot10^4$) at all photon numbers, compared with the other investigated Ta films. The (poor) RF performance of these beta-Ta rich thin-films seems to be dominated by power-independent losses rather than by power-dependent TLS loss. Contrary to this, the alpha-Ta rich films show a photon number-, i.e. power-, dependent behavior. In particular, they have significantly higher $Q_i$ values at $n\thickapprox1$: Ta on TiN has a $Q_{i,{TiN+Ta}}$ = $0.91\cdot10^6$, Ta sputtered at 600°C on Si has $Q_{i,{600^{\circ}C}}=1.27\cdot10^6$ and Ta on Nb has a $Q_{i,{Nb+Ta}}$ = $1.46\cdot10^6$. Intriguingly however, the resonators from Ta on TaN represent a notable exception: they perform significantly worse with only $Q_{i,{TaN+Ta}}\approx0.13\cdot10^6$. The measurements were reproduced on a separate batch of chips, yielding similar results (see Supplementary Information, Fig.~S4). This finding is remarkable, because in particular all Ta films on the three different seed layers had practically identical values in terms of resistivity, critical temperature, RRR and XRD peak positions. This indicates that while the Ta grown on these seed layers appears to be of equal composition - namely predominantly comprising the alpha-phase, this alone is apparently not sufficient to guarantee a high RF performance. Rather, for the alpha-Ta on TaN resonators, there must be another limiting factor, apart from the Ta film material itself. Suspecting the origin of the responsible losses to be situated in the seed layer material and/or at the involved interfaces (Si-TaN and/or TaN-Ta), we set out to specifically investigate their potential role, as will be substantiated in the remainder of the paper. For these studies, we focused on those material stacks comprising seed layers that gave best (Nb) and worst (TaN) performing resonators. At first, by systematically varying the thickness of the seed layers, we investigated whether losses in the 'bulk' of the two seeds would play a limiting role. Subsequently, we analyzed the data from complementary XRD and high-resolution TEM imaging to differentiate the possible impact of different amounts of built-in microstrain and structural defects in these two exemplary thin-film stacks.


\subsection{Influence of seed layers thickness on Ta thin film RF-performance}

To investigate if the different RF-performances of the Ta thin films on different seed layers shown in Fig.~\ref{Ta_RF} is due to intrinsic losses from these layers themselves, a study with varying Nb and TaN seed layer thicknesses was conducted. These seed layers were selected because the Ta on them has the highest (Ta on Nb) and the lowest (Ta on TaN) quality factors as discussed previously. The substrate and the cleaning processes are the same as outlined earlier. After the Si substrates were prepared and loaded into the sputtering chamber, Nb or TaN was deposited with different thicknesses. The thickness of the seed layers was determined by AFM measurements on partially coated test chips, as explained in Section IV (Methods). The thinnest Nb and TaN layers were 3~nm, followed by 10~nm, 15~nm and 20~nm. Because the AFM measurements are done after the chips are taken out of the vacuum, it is highly likely that an oxide layer has already formed. So the thickness of the seedlayers should be treated with a $\pm$1~nm tolerance to account for that. This is necessary to distinguish from the full stacks including Ta, because for those films, after each Nb or TaN deposition, the vacuum is not broken and 200~nm Ta is deposited in-situ on top of the seed layers. 

\begin{figure*}[!t]
  \includegraphics[width=\textwidth]{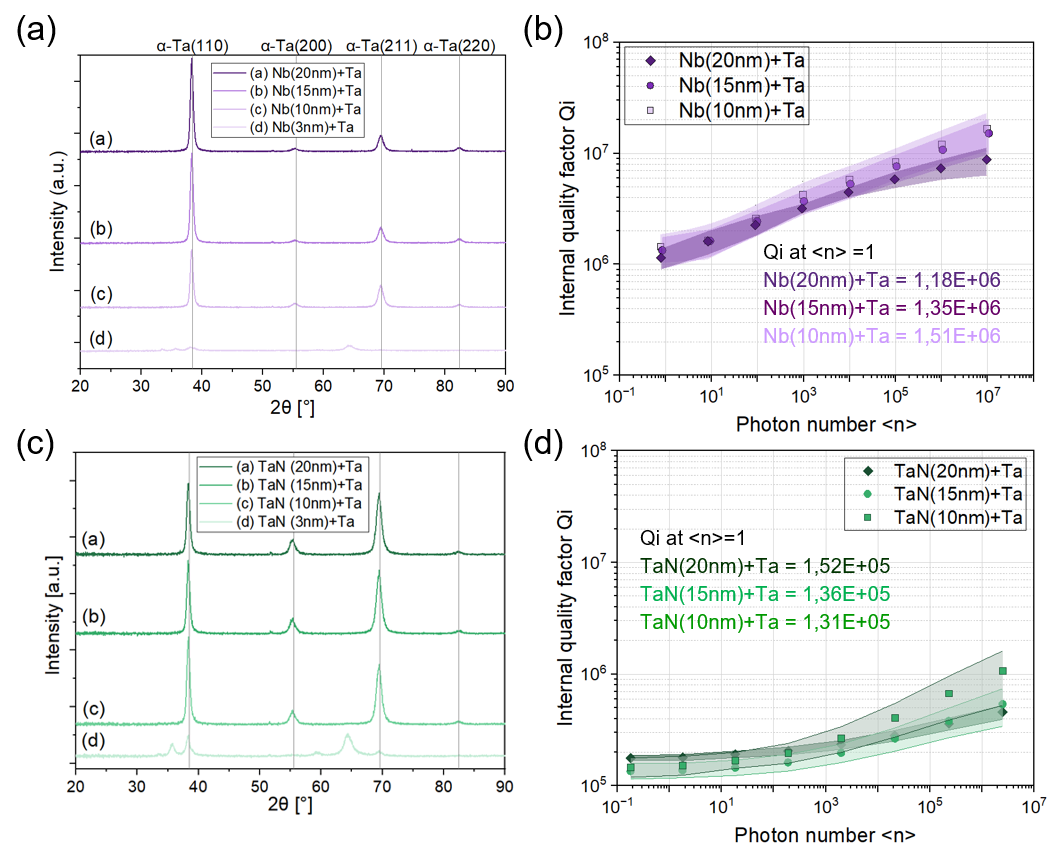}
  \caption{Tantalum thin films with a thickness of 200~nm sputtered on Nb and TaN seed layers of different thickness. (a) GI-XRD spectra for Ta on Nb showing that a 3~nm thick Nb seed layer is not thick enough to nucleate alpha-Ta growth. With thicker seed layers the alpha-Ta growth works and only peaks at alpha-Ta positions are present. (b) Internal quality factors of CPW resonators fabricated from the chips with different Nb seed layer thicknesses where the alpha-Ta growth was successful. The mean $Q_i$ value slightly decreases with increasing seed layer thickness in single-photon regime. But taking error margins into account (95 $\%$-confidence interval, compare Fig. 4), the performance of the thin films is comparable.(c) GI-XRD spectra for Ta on TaN with different thicknesses. Here, the same can be observed: a 3~nm thick layer does not nucleate alpha-Ta growth, while thicker layers do. (d) Resonator measurements of alpha-Ta on different TaN thicknesses. The opposite trend as in the case of Nb, namely here an increasing $Q_i$ with thicker seed layer, can be observed in the single-photon regime. But, as before, taking the error margins into account, the performance is not dominated by the seed layer thickness.}
  \label{SeedThickness}
\end{figure*}

The obtained Ta thin films on Nb seed layers with different thicknesses were investigated with GI-XRD to analyze their crystallographic composition, see Fig.~\ref{SeedThickness}(a). The top curve (a) corresponds to the thickest seed layer with 20~nm thickness. The spectrum has only peaks at positions where alpha-Ta is expected, indicating that the Ta that grows on top of this seed layer is purely alpha, which is in line with previous results \cite{urade_microwave_2024}. The spectra for Ta on 15~nm and 10~nm thick Nb seed layers look alike in terms of peaks present. In contrast, the Ta sputtered on the 3~nm thick Nb seed layer shows a spectrum more comparable to the Ta sputtered at 20°C directly on Si,  cf. Fig.~\ref{XRD}. Here, we see also beta-Ta peaks (002) and (321) as well as (403), while the only remaining alpha-Ta (110) peak is just weakly present. This might suggest, that the 3~nm thick Nb layer is not thick enough to nucleate the alpha-Ta growth, potentially due to pinholes/non-uniformity in the layer. Another reason could be that the first few atomic layers of Nb that grow on the Si, are adapting to the Si crystal lattice and only after a few layers the Nb can grow in its preferred body-centered-cubic structure, which is needed to facilitate body-centered-cubic alpha-Ta growth. 

\begin{figure*}[!t]
  \includegraphics[width=\textwidth]{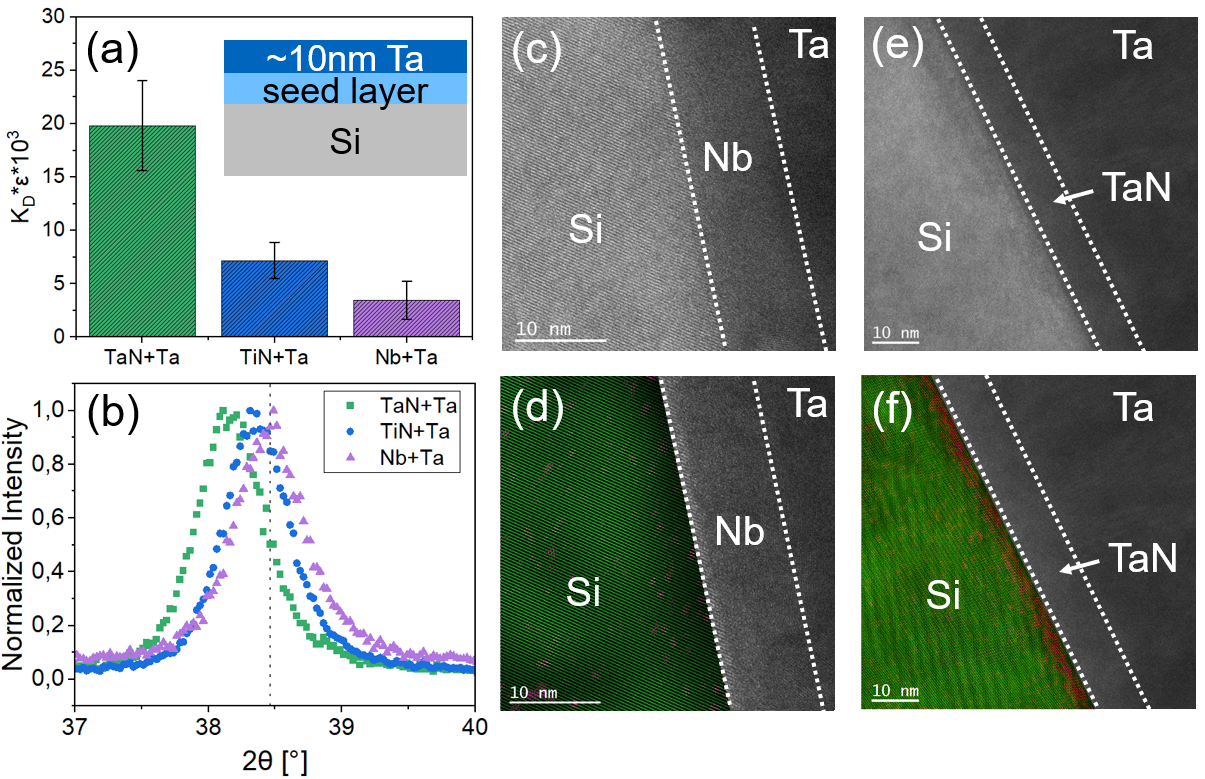}
  \caption{Strain analysis at the interfaces of the thin films. (a) Results for $K_D\cdot \varepsilon$ from Williamson-Hall analysis of the diffraction spectra for circa 10~nm thin Ta films on the three different seed layers. Ta on the TaN seed layers has the highest value, while on TiN it is significantly lower, and on Nb the lowest value is obtained. This correlates inversely with the internal quality factors displayed in Fig.~\ref{Ta_RF}. (b) Zoomed-in diffraction peaks of alpha-Ta (110) from the same measurement. The dashed vertical line serves as reference of the peak position of relaxed tantalum \cite{LEE2004Texturestructurephasetransformationsputterbetatantalumcoating}. The individual shifts in peak position are consistent with the trend observed from the microstrain analysis, here indicating higher global (macroscopic) strain in the Ta on TaN and lowest strain in Ta on Nb. (c) HRTEM image of Ta on Nb. The interface looks smooth, which is also confirmed by the IFFT analysis (d) where a uniform lattice spacing (green) in the Si can be seen all the way up to the interface with Nb. (e) HRTEM image of Ta on TaN where a distorted line can be seen along the interface within Si. The IFFT-image (f) confirms this, showing in pink the regions where Si atoms are spaced differently than in the unstrained bulk region (green). Note that the undisturbed, bulk Si lattice spacing (further away from the interface) was used as reference. Therefore, the overlaid, colored IFFT-images are displayed only within Si. Smaller pink spots away from the interface (present in both images) are assigned to IFFT artifacts. 
  }
  \label{TEM}
\end{figure*}

The Ta on Nb thin films for which the nucleation of alpha-Ta growth seemed to have been successful were then patterned into CPW resonators, each fabricated with the same method and measured with the same set-up and under the same conditions as mentioned earlier. The resulting $Q_i$ values are shown in Fig.~\ref{SeedThickness}(b) on the right. The Ta thin film on 10~nm thick Nb has a single-photon $Q_{i,Nb10nm}$ of $1.51\cdot10^6$, which is similar to the previously reported value for the same film, see Fig.~\ref{Ta_RF}. The Ta film on 15~nm of Nb has a $Q_{i,Nb15nm}$ of $1.35\cdot10^6$ and the one on 20~nm Nb a $Q_{i,Nb20nm}$~=~$1.18\cdot10^6$. 

A corresponding analysis was performed for the Ta on TaN seed layer films. The GI-XRD spectra in Fig.~\ref{SeedThickness}(c) show a similar seed layer thickness dependence as before in (a), for Ta on Nb: The 3~nm seed layer is not thick enough to nucleate alpha-Ta growth on top and the spectrum looks mixed with beta-Ta peaks. For the thicker seed layers (10, 15 and 20~nm) the alpha-Ta growth does work. The internal quality factor results for the three alpha-Ta films are shown in (d). Here, we record values of $Q_{i,TaN10nm}$ of $1.31\cdot10^5$, $Q_{i,TaN15nm}$ of $1.36\cdot10^5$ and $Q_{i,TaN20nm}$ of $1.52\cdot10^5$. 

Comparing both results for quality factor change with seed layer thickness, an opposite, though weak trend for Ta on Nb (where $Q_i$ decreases by about 20\%) with respect to Ta on TaN  (increase by about 15\%) can be noted. For Nb, this decrease might originate from the increased thickness of lossy Nb-oxide, formed at the exposed resonator edges. For TaN, starting from a likely poor interface with structural defects (as will be further outlined below) a possibly higher quality film may gradually grow with increasing thickness. Such effect might not be observable for the Nb seed, whose interface is potentially already optimized.

As however, all recorded maximum $Q_i$ changes with seed layer thickness are well below a hypothetical 100\% change (anticipated for a pure 'bulk' effect, when changing the layer thickness from 10 to 20 nm) and, in addition, still within the error ranges of the measurement (see,  Fig.~\ref{SeedThickness}), we here conclude that intrinsic losses from the seed layers are not dictating the RF-performance in either of the two cases (Nb or TaN). Rather, this finding suggests that the dominating TLS losses are originating from the various involved interfaces between substrate, seed and superconductor (recall that losses from the Ta surface oxides are supposedly the same for all samples, due to same surface treatment steps). 

We therefore carried out a detailed inspection of these interfacial regions, focusing on both, direct signatures for strain (in the crystallographic spectra), and visible lattice deformations (in TEM analysis).


\subsection{Strain in Ta thin films}


Strain is known to couple directly to TLSs, thereby modifying their RF characteristics.  \cite{CARRUZZO2021Distributiontwolevelsystemcouplingsstrainelectricfieldsglasseslowtemperatures, LISENFELD2015Observationdirectlyinteractingcoherenttwolevelsystemsamorphousmaterial, lisenfeld_electric_2019, schechter_inversion_2013}. Since the described losses in our Ta resonators can be well assigned to TLSs residing at the interfaces, strongly linked to material and structural defects, and local disorder in this area, the determination of strain in the thin films is a most helpful tool to narrow down the different loss characteristics.

Therefore, we conducted a systematic strain analysis for the Ta on Nb, TiN and TaN seed layer material stacks. Here, in order to focus on interfacial contributions, we sputter deposited only 10~nm of Ta on top of the seed layer materials using the same chemical cleaning and sputtering procedure as before. By that, we could specifically probe the interface-near region with GI-XRD, which would otherwise (for 200 nm Ta) not be well accessible due to significant X-ray attenuation and superimposed signals from the bulk-Ta layer. 

The recorded diffraction spectra of Ta on Nb, TiN and TaN were analyzed in terms of peak position 2$\theta$ and Full-Width-Half-Maximum (FWHM)~$\beta$ (for details see Supplementary Information: Lattice constant and strain calculation from XRD-data). By extracting these two parameters for several peaks in the spectra and carrying out a Williamson–Hall analysis, the built-in microstrain $\varepsilon$ and the grain size \textit{D} of the films are simultaneously obtained from fits to the data (for details, see section IV. Methods). 

The extracted microstrain values for all three different films are displayed in Fig.~\ref{TEM}(a) (note that, we display the product $K_D\cdot \varepsilon$, where $K_D$ is a model-dependent constant scaling factor in the range 1-2, see section IV. Methods). The film with the highest average microstrain is Ta on TaN ($K_D\cdot\varepsilon_{TaN}$~=~19.8~$\cdot10^{-3}~\pm~4.2\cdot10^{-3}$), followed by Ta on TiN ($K_D\cdot\varepsilon_{TiN}$~=~7.2~$\cdot10^{-3}~\pm~1.7\cdot10^{-3}$), and finally Ta on Nb with the lowest value ($K_D\cdot\varepsilon_{Nb}$~=~3.5~$\cdot10^{-3}~\pm~1.8\cdot10^{-3}$).

The absolute strain values are subject to relatively large error margins (see, section IV. Methods). Notwithstanding that, the overall trend of $\varepsilon$ (Ta on TaN) > $\varepsilon$ (Ta on TiN) >  (Ta on Nb) remains obvious and correlates remarkably well with the resonator quality factors reported before - namely just inversely: highest $Q_i$ $\leftrightarrow$ smallest $\varepsilon$ .  Extracted grain sizes were in the range of 15 to 65 nm, see Supplementary Information for detailed values. 

While microstrain generally reflects a microscopic fluctuation of the lattice constant, thereby resulting in (angle dependent) XRD peak broadening, a measurable shift of the peak position would relate to a global, uniform change of the lattice constant and thereby signify macroscopic (tensile or compressive) strain. Fig.~\ref{TEM}(b) shows the alpha-Ta (110) peaks in the spectrum of all three thin films under study: the peak of the Ta on Nb is close to the reference position of relaxed bulk alpha-Ta, while the one for Ta on TiN is slightly shifted towards smaller $2\theta$ values, and the one for Ta on TaN is shifted even more, corresponding to the strongest (relative) increase of lattice constant in these latter samples, possibly caused by tensile strain in the direction of X-ray diffraction. This trend is consistent with the above reported microstrain trend, both indicating lowest strain for samples with highest quality factors, and vice versa. 


Where does the strain in the thin film interfacial regions come from, and why is it different for different material combinations? One parameter that should be taken into consideration, is the different lattice constants of the materials that are brought in contact with each other. At the interface of two materials, a lattice mismatch commonly induces stress, which, depending on the elastic constants of the materials, transforms into a measurable deformation (strain). If, for increasing thickness, the stress relaxes into the formation of dislocations or other defects such as microcracks, further loss mechanisms (non-thermal phonon bursts) may result \cite{ANTHONY-PETERSEN2024stressinducedsourcephononburstsquasiparticlepoisoning}. 
A basic comparison of the lattice constants of the materials involved in our study gives already important insight: the lattice constants of Ta on Nb are almost similar as shown in Tab. \ref{t1}, which correlates well with the lowest measured strain in our analysis. 

\begin{table}[ht]
\centering
\caption{Lattice constants of alpha-Ta and seed layer materials\cite{HALLMANN2013Effectsputteringparameterssubstratecompositionstructuretantalumthinfilms, HULPKE1992Latticedynamicsniobium001surface, IGASAKI1978StructureElectricalPropertiesTitaniumNitrideFilms, TERAO1971StructureTantalumNitridesa}.}
\begin{tabular}[t]{cccc}
\hline
Material&Crystal&Lattice constant \\
&structure&[Å]\\
\hline
$\alpha$-Ta &bcc &3.306\\
 Nb& bcc &3.300\\
TiN& fcc &4.241\\
TaN& fcc &4.336\\
\hline
\end{tabular}
\label{t1}
\end{table}%

In contrast, TiN has a larger lattice constant than Ta, which creates a significant mismatch to alpha-Ta. This mismatch is even more pronounced for cubic TaN with respect to Ta. These comparisons are based on the assumption that face-centered cubic (fcc) tantalum nitride and titanium nitride are present, as these are the only phases we observe in XRD (see Supplementary Information, Fig.~S5). The fact that alpha-Ta has a body-centered cubic (bcc) crystal structure is anticipated to add further to the lattice incompatibility. 

In addition to these likely responsible lattice mismatch effects, the physical process of film deposition itself can well contribute additionally. It is established that stress can be incorporated in sputtered thin films \cite{windischmann_intrinsic_1992, thornton_stress-related_1989, freund_thin_nodate}. Especially in reactively sputtered nitrides, effects like atomic peening are common, where nitrogen would be incorporated interstitially, leading to low atom mobility in the nitrides during growth, which creates higher-density layers featuring built-in stress. Such effects strongly depend on the particularly chosen sputter deposition parameters. For example, a trend similar to our observation has been reported for TiN as resonator material, where higher deposition pressures resulting in reduced film stress correlate with enhanced internal quality factors \cite{OHYA2013SputteredTiNfilmssuperconductingcoplanarwaveguideresonators}.


Finally, again dependent on material choice and deposition parameters, also the supporting silicon substrate may be affected from the sputter deposition process, which - especially for reactive sputtering - involves considerable ion bombardment of the surface. In order to investigate possible structural defects in the interface-near silicon substrate, we carried out a detailed High-Resolution TEM (HRTEM) analysis, thereby focusing on those both thin film stacks for which we recorded the highest (Ta on TaN) and lowest (Ta on Nb) strain values. 

As visible in Figures 6(c) and 6(e), the Nb/Si interface exhibits a sharp and well-defined transition between the two materials. In contrast, the TaN/Si interface is characterized by a narrow interfacial region extending several nanometers into the Si substrate, where the image contrast differs from that of the surrounding Si lattice. 
To assess whether the observed contrast deviation of the interfacial region can be assigned to the local lattice periodicity that differs from that of the bulk Si crystal, (inverse) fast Fourier transform (FFT/IFFT) analysis was performed using a defect-free Si region as a reference (see Supplementary Information for details). The resulting reconstructions are shown in Figures 6(d) and 6(f). The green overlay corresponds to the periodicity of the reference Si lattice, whereas the pink contrast highlights residual frequency components that deviate from the reference periodicity. For the Nb/Si interface (Figure 6(d)), only a limited contribution from residual frequency components is observed. In contrast, the TaN/Si interface (Figure 6(f)) exhibits a pronounced localization of these components within the Si region adjacent to the interface. This observation indicates the presence of a structurally perturbed region at the TaN/Si boundary, consistent with localized lattice distortion and strain accommodation within the Si substrate.

 Our HRTEM-based analysis provides local structural information with nanometer-scale spatial resolution and is particularly sensitive to the single-crystalline Si substrate, which serves as a well-defined crystallographic reference. In contrast, owing to the nanocrystalline nature of TaN, an equivalent local strain analysis within this film is not readily achievable. Still however, the localized distortion detected in Si likely reflects strain accommodation at the TaN/Si heterointerface arising from the overall stress state of the Ta/TaN stack. The combined observations suggest that strain is distributed across the multilayer structure and is partially transferred to the underlying Si, where it can be directly visualized as a confined interfacial strain field.


A possible further explanation for the apparent, strong lattice
distortions is unintended, shallow implantation and defect
creation by ion bombardment during reactive sputtering.
Alternatively, shallow inter-diffusion of nitrogen into the Si
and a resulting silicon nitridation may have occured. We
performed an EELS analysis (see Supplementary Fig. S8),
with clear identification of the stack-comprising elements but
no evidence for nitrogen incorporation into Si - likely because
of insufficient resolution for such anticipated, very shallow
interfacial layers.

Such sputter-deposition related silicon modifications, potentially even resulting in nanometer-thin amorphous layers, are likely sources of TLSs and related losses \cite{urade_microwave_2024}. At the same time, they would also likewise contribute to strain in this interfacial region. As however, they are in general common to both (similar) reactive sputter depositions, for TiN and TaN, one main reason for their dissimilar RF performance must be originated in the structural properties and chemistry at the interface. Indeed, TaN as a compound comprising Ta (the material of the layer above), is prone to form intermixing layers and graded interfaces \cite{zaman_microstructure_2017}, which may result in defective transition layers where TLSs would reside due to atomic defects and dangling bonds. TiN on the other hand, exhibits improved crystalline quality and reproducible growth on Si. In particular, TiN has been demonstrated to grow epitaxially and reproducibly on silicon substrates, while TaN often requires a TiN buffer layer to achieve high crystalline quality \cite{zhang_microstructure_2013,wang_epitaxial_2002}. 

We hypothesize that the distortions from the deposition process of TaN on Si create strain at the metal-substrate interface. Together with the micro and macroscopic strain in the Ta to seed layer interface, due to lattice mismatching, significantly different strain situations are present in these thin film stacks. Given the rather stiff, rigid material TaN (Young modulus ca. 250 GPa), any present stress may be transferred in both directions in the stack (top-down, and bottom-up, cf. Fig. 1 c). This contrasts to Nb as a seed layer, which is both more ductile (Young modulus ca. 105 GPa) and almost perfectly lattice-matched to Ta.  We assign the strikingly different RF performance to originate from the described different lattice distortions and structural defects at the seed layer-substrate and seed layer-Ta interfaces, which in turn host different TLS distributions. At the same time, these material imperfections together with the lattice mismatches are sources of strain, which likewise affects the TLSs and the losses originating from them. A potentially complex combination and interplay of these contributions is most likely responsible for the observed, clear correlation of strain and resonator quality factor.


\section{DISCUSSION}

The results presented in our study establish a clear connection between interfacial microstructure, microstrain and superconducting microwave resonator performance of sputtered tantalum thin films. While previous studies had already suggested that the alpha-phase of tantalum is preferable for achieving low-loss qubit devices, our data demonstrate that this phase purity alone does not guarantee high resonator performance. In particular, alpha-Ta films with comparable resistivity, critical temperature, and crystallographic properties exhibit markedly different internal quality factors depending on the underlying seed layer, indicating that additional mechanisms beyond bulk material and metal-to-air interface losses would govern microwave losses.

A central finding of this work is the strong correlation between interfacial strain and resonator performance. Williamson–Hall analysis revealed a systematic decrease in microstrain from Ta on TaN via Ta on TiN to Ta on Nb seed layers, which directly correlates with the measured internal loss of the resonators, decreasing in the same order. These strain analysis studies are complemented by HRTEM analysis, which directly visualizes significant interfacial disorder in the case of TaN-based stacks - particularly in the interface-near, topmost Si region. 

We take these observations as strong evidence that strain and lattice distortions at interfaces play a critical role in determining RF losses.

The physical origin of these losses can be explained by TLSs, which are widely accepted as a dominant loss mechanism in superconducting resonators at low temperatures and low photon numbers. Strain is known to modify the local atomic configuration, potentially creating more, or activating already existing TLS. Microscopically, these  form at atomic configurations, where single or multiple atoms can tunnel between different sites. Dangling bonds, lattice distortions, or defect complexes play a significant role in their characteristics. In our case, the increased strain observed for Ta on TaN is accompanied by a distorted interfacial region in the silicon substrate, as revealed by TEM and IFFT analysis. Such disordered interfacial layers are expected to host a higher density of TLSs, which, via their associated, atomic dipole moments, may couple to the electric field of resonators and give rise to enhanced dielectric losses. Conversely, the near perfectly lattice-matched Ta on Nb stack exhibits minimal strain and a structurally sharp, undisturbed interface, consistent with the highest observed quality factors.

Our study on the variation of seed layer thickness provided further insight into the origin of these losses. Despite doubling the thickness of the Nb and TaN seed layers, no such significant change in resonator performance is observed. This indicates that intrinsic dielectric losses within the seed layer itself are not the dominant limitation. Instead, also these results support our conclusion that the interfacial regions, specifically the substrate to seed layer and seed layer to Ta ones, are the primary locations of loss. This interpretation is consistent with prior studies identifying the interface to the substrate as one dominant host region for TLSs in superconducting circuits,  \cite{WENNER2011Surfacelosssimulationssuperconductingcoplanarwaveguideresonators} and further highlights that buried interfaces, even when not directly exposed, can critically impact device performance.

The markedly reduced performance of Ta on TaN compared to Ta/Nb and Ta/TiN, despite similar crystal structure and DC electrical properties, suggests that chemical and structural effects specific to reactively sputtered nitrides play an additional role. Compared to TiN, TaN has been reported to form intermixed and defective interfaces, which can enhance disorder and facilitate TLS formation. In particular, different effects caused by nitrogen ions (shallow diffusion, nitridation) may have led to the visibly distorted Si interfacial region. In addition, film growth- and interface chemistry-related effects, specifically for TaN on Si, are likely to have contributed to these observations. We assign these contributions as responsible for the measured differences between TiN and TaN seeds in terms of RF performance. 


In conclusion, we suggest that the intriguing correlation of strain and lattice distortions with RF performance results from a complex combination and interplay of these with the TLSs that are responsible for the apparent dielectric losses: structural material imperfections and lattice mismatches are hosting an increased number of TLSs and at the same time causing interfacial strain. The strain itself likewise affects and potentially enhances the TLSs' loss characteristics. Here, the bi-directional stress through the seed layer is likely differently transferred, depending on the material's rigidity (TaN > Nb).

Beyond the specific material system studied here, our findings have broader implications for superconducting quantum device fabrication. They suggest that optimizing thin film performance requires not only control over phase composition and bulk film properties, but also careful engineering of interfaces - in particular of those to the substrate - to minimize strain and disorder. Selecting seed layers with low lattice mismatch,  stable interface chemistry and low-defect growth characteristics lends itself as a key strategy to effectively suppress TLS related losses. This perspective may also help to understand previously reported variations in resonator and qubit performance across different material stacks and fabrication processes, beyond Ta as superconducting thin-film material.  Future work combining additional surface analytical methodologies such as advanced diffraction and nanoscale elemental identification techniques with atomistic modeling may provide a more detailed understanding of the interplay between strain, disorder, and TLS formation at interfaces to superconductors.

\section{Methods}

\subsection{Sputter deposition of thin films}

The investigated tantalum thin films were deposited with an RF-magnetron sputtering tool (Alcatel). At first, high-resistivity (>10 kOhm$\cdot$cm) silicon (100) substrates were diced into 10 mm x 6 mm chips. The silicon chips were cleaned by ultrasonication in acetone and isopropyl alcohol for three minutes each and subsequent rinsing in the respective solvent, followed by immersing the chips into buffered oxide etchant (BOE,~7:1~HF:NH\textsubscript{4}F) for 60 seconds, in order to remove the native oxide on the silicon [\textit{Caution: hydrofluoric acid (HF(aq)) is very hazardous to health; special care/training is mandatory}]. After these chemical cleaning steps, the substrates were rinsed with DI-water and dried in a flow of nitrogen. 

After these steps, the cleaned substrates were loaded into the sputtering tool and the chamber was evacuated to a base pressure of $\sim10^{-7}$mbar. The sputter target-to-chip distance was set to 7~cm, as process gas Argon (Ar) was used with a flow rate of 20~sccm, and 18~$\mu$bar pressure and a power of 200~W were used for sputtering. We followed two different routes for thin-film deposition: in the first case, we sputter-deposited ca. 200~nm of Ta directly on silicon at different temperatures ranging from room temperature (20°C) to 600°C, using a heated substrate holder. In the second case, we at first deposited 10~nm thin seed layers of different materials, followed by 200~nm of Ta, all at room temperature. In case of the Ta thin films with seed layers, the respective target (Ti, Nb) was mounted in the sputtering chamber, together with the Ta target. A 2~sccm flow of nitrogen was added during the reactive sputtering process to obtain TaN and TiN seed layers. After the deposition of a given seed layer, the chips remained under vacuum in the sputtering chamber, and they were rotated towards the tantalum target from where the subsequent tantalum deposition was carried out. 

The thickness and sputter rate for each of the materials was determined by sputtering on photoresist-patterned silicon reference chips followed by lift-off. The step height of the films was subsequently measured with a profilometer or AFM, and the sputter time was optimized to obtain Ta films with a thickness of 200~nm and seed layers with a thickness of 10~nm. We note that, the actual height of the seed layers below the Ta films might deviate slightly from the determined height of the seed layers, because for their height measurements, the thin films have to be taken out of the vacuum directly after deposition such that oxidation cannot be avoided. In contrast, for the Ta thin films on top of seed layers, the vacuum is not broken and Ta is deposited in-situ on them, preventing any prior native oxide formation at the interface. However, we still anticipate that the height measurements give a solid estimate for the sputter rates and needed sputter times. 

\subsection{Surface roughness measurements}

Within a few hours following deposition, the surfaces of the planar tantalum thin films were investigated with an AFM (Bruker, Dimension 3000). The AFM was operated in tapping mode and measurements with a scan area of 5~$\mu$m x 5~$\mu$m, at a scan rate of 1~Hz and sampling rate of 512 data points per line were performed. Subsequently, the root mean square roughness values of the films were extracted from the obtained AFM images using Gwyddion software. 

\subsection{Crystallographic characterization}

For crystallographic characterization, XRD measurements were carried out on the planar thin-film surfaces. A GI-XRD capable machine (Panalytical, Empyrean) was used for the measurements. In GI-XRD, a small incidence angle of the X-ray source is kept constant which minimizes the background signal and maximizes the signal that comes from diffraction at the investigated surface. The measurements were done with a fixed angle of 0.5° of the incidence beam with respect to the surface plane. A scanning range of $2\theta=$ 20° to 90° was covered by the moving receiver. The obtained diffraction spectra were fitted with a Gaussian and the 2$\theta$ position of peaks and full width at half maximum (FWHM) $\beta$ values were extracted. Values were corrected for the instrument broadening, which was determined to be $0.304^\circ \pm 0.011^\circ$ in this set-up (see Supplementary information for more information). With these parameters, the Williamson-Hall equation\cite{KHORSANDZAK2011XrayanalysisZnOnanoparticlesWilliamsonHallsizestrainplotmethods, MAEvaluationlatticestrainZnOthinfilmsbasedWilliamsonHallanalysis,BIRKHOLZThinFilmAnalysisXRayScattering}

\begin{equation}
\beta\cdot\cos(\theta)=\frac{K_L\lambda}{D}+K_D2\varepsilon\sin(\theta)
\end{equation}

was used to extract both the crystallite size \textit{D} and the microstrain $\varepsilon$ based on full Williamson-Hall analysis. This allowed for a later comparison of these structural parameters to the RF performance of the thin films in terms of resonator quality factors. The Scherrer constant $K_L$ is assumed to be 0.9 and $K_D$ is a scaling factor which is correlated to the underlying lattice distortions and the actual strain model used \cite{BIRKHOLZThinFilmAnalysisXRayScattering}. In order to keep our analysis independent from  assumptions at this point, we leave $K_D$ in the equation and rather extract the product $K_D\cdot\varepsilon$, which readily allows for relative comparison (for details, see the Supplementary Information). 

We note that only a limited number of peaks (typically four) in the spectra could be used for analysis. For that reason, the extracted values of strain and crystallite size are subject to significant error margins. But as the three investigated spectra (which are shown in Supplementary Fig.~S6) are all from alpha-Ta and the same four alpha-Ta peaks could be used for every fit, a solid relative comparison between the parameter sets was well possible. 

\subsection{Electrical measurements of sputtered thin films}

For electrical characterization, the room-temperature resistivity and temperature dependence of the resistance were measured, to determine the residual resistance ratio (RRR) and superconducting critical temperature. For this purpose, a bar-shaped geometry (300~$\mu$m width x 3600~$\mu$m length) was patterned from a given Ta thin-film, using photolithography and Reactive Ion Etching (RIE, Oxford, PlasmaPro 80 Cobra) with CF\textsubscript{4} gas (25~sccm, 30~mTorr, 70~W HF-power, 150~W ICP-power, 150s etch time), featuring voltage-probe leads at either side of the bar. By forcing a DC constant current of 100 $\mu$A through the bar and measuring the voltage drop between two adjacent leads (Keithley, SMU 2635A/B), four-point resistance measurements were performed. Using Ohm's law together with the known dimensions of the structure, the room-temperature resistivities of the Ta films were determined. 

The same structure and measurement setup was used to measure the critical temperatures $T_c$ of the thin films (transition to superconducting state). For these measurements, the chips were wire-bonded and mounted into a cryostat (kiutra, L-type Rapid). The temperature was swept over a range of 400~mK to 7~K and the resistance was continuously measured. In each measurement, one temperature up and one temperature down sweep was performed, in order to check if either transition temperatures differ, which could indicate that the sample has a delay in thermalization. However, this was not the case and the transitions for both, the up-sweep and the down-sweep, were always lining up on top of each other. From the room temperature (RT) resistance and the resistance shortly before the superconducting transition, the RRRs were calculated, which is the quotient of their resistance values: $RRR=R_{RT}/R_{above T_C}$.  

\subsection{Resonator measurements}

The suitability of the Ta thin films towards their application in superconducting qubits was evaluated by fabricating coplanar waveguide (CPW) resonators. The losses in a system are characterized by the loaded quality factor $Q_l$

\begin{equation}
Q_l=\omega\cdot\frac{energy\ stored\ in\ resonator}{energy\ loss\ per\ second}
\end{equation}

where $\omega$ is the angular frequency. $Q_l$ is comprising two contributing portions of energy loss which are characterized by the internal quality factor $Q_{i}$ for the intrinsic losses and the coupling quality factor $Q_{c}$, which is a design parameter and is determined by the geometry of the resonator which couples to the feed line \cite{ZMUIDZINAS2012SuperconductingMicroresonatorsPhysicsApplications}. The three quality factors are connected via the following relation \cite{BESEDIN2018Qualityfactortransmissionlinecoupledcoplanarwaveguideresonator}

\begin{equation}
\frac{1}{Q_l}=\frac{1}{Q_i}+\frac{1}{Q_c}.
\end{equation}

A design with twelve resonators in hanger geometry, which are capacitively coupled to a common feed line (see Supplementary Fig. S9), was created and patterned into the Ta by photolithography and the same CF\textsubscript{4} gas RIE recipe mentioned before. The resonators are designed to operate in the over-coupled regime, meaning that $Q_{i}>Q_{c}$. They feature all different resonance frequencies in the range 5-9 GHz. Around the resonators, holes are patterned in the ground plate to trap magnetic vortices \cite{CHIARO2016Dielectricsurfacelosssuperconductingresonatorsfluxtrappingholes}. The etching recipe was optimized to over-etch the Ta ca. 20~nm into the silicon to guarantee that the superconducting film is fully etched and to increase resonator performance by trenching \cite{CALUSINE2018Analysismitigationinterfacelossestrenchedsuperconductingcoplanarwaveguideresonators, WENNER2011Surfacelosssimulationssuperconductingcoplanarwaveguideresonators}. After the etching and stripping of the resist, the chips were immersed in BOE for 10 minutes in order to remove the native oxides on the Si and Ta surfaces and possible defects from RIE etching. After that, the chips were bonded to a PCB via aluminum wire bonds. The edges of a certain chip were also contacted to the PCB via bond wires for grounding, and additional bond wires across the feed line were placed to mitigate unwanted electromagnetic field modes and to reduce potential differences between different areas of the ground plate. The chip and the PCB were then placed in a copper box which served as electromagnetic shield while also ensuring proper thermalization. Two SMA connectors were finally used to connect the ends of the feedline on the PCB to the wiring in the cryostat (see Supplementary Fig. S10 for a detailed description of cryostat wiring).

A vector network analyzer (VNA), model Keysight P5002B, was used to probe the resonators with RF-signals by performing S21-measurements.  First, a coarse frequency sweep over the range of 4-9~GHz was performed and the resonance peaks were detected. After this, a finer frequency sweep with a narrower frequency range around the found resonances was carried out, with a power ranging from -90~dBm to -160~dBm in -10~dBm steps. The transmitted signal was detected and normalized with the input signal from the VNA to obtain the scattering parameter S\textsubscript{21}. The transmission signal of a capacitively coupled resonator in hanger geometry is described by \cite{BaityPhysRevResearch}

\begin{equation}
S_{21}=ae^{i\alpha-2i\pi\cdot f\tau}\left[1-\frac{(Q_l/|Q_c|)e^{i\phi}}{1+2iQ_l(f/f_r-1)}\right]
\end{equation}

in the frequency domain, where $f$ is the signal frequency, $f_r$ is the resonance frequency and $\phi$ is the resonance phase shift due to impedance mismatches. The term in brackets describes an ideal resonator, but in reality there are environmental distortions which are taken into account by the pre-factor. There, $a$ describes changes in the signal amplitude due to cable dampening, $\alpha$ accounts for possible phase offsets and $\tau$ for a delay due to the length of the cables. In the complex plane, the $S_{21}$ response is a circle with diameter $Q_l/|Q_c|$. By applying a circle fit \cite{ DENG2013analysismethodtransmissionmeasurementssuperconductingresonatorsapplicationsquantumregimedielectriclossmeasurements, KHALIL2012analysismethodasymmetricresonatortransmissionappliedsuperconductingdevices, PROBST2015Efficientrobustanalysiscomplexscatteringdatanoisemicrowaveresonators, MEGRANT2012Planarsuperconductingresonatorsinternalqualityfactorsonemillion}, the loaded and coupling quality factors can be extracted and the internal quality factor can be determined (see Supplementary Fig. S10 for an example circle fit). In order to evaluate the RF-performance of the resonators as function of average photon number <\textit{n}> in the resonator, thereby in particular addressing the single-photon regime, each measurement RF power was converted into <\textit{n}> using

\begin{equation}
<n>=\frac{1}{\pi\cdot h \cdot(f_r)^2}\cdot\frac{(Q_l)^2}{Q_c}\cdot\frac{Z_0}{Z_r}\cdot P
\end{equation}

where $f_r$ is the resonance frequency, $Z_0$ is the coupling impedance, $Z_r$ is the impedance of the resonator and $P$ is the applied power \cite{MCRAE2020Materialslossmeasurementsusingsuperconductingmicrowaveresonators,BRUNO2015Reducingintrinsiclosssuperconductingresonatorssurfacetreatmentdeepetchingsiliconsubstrates}.

\subsection{Interface and strain analysis}


Transmission electron microscopy (TEM) was performed on selected thin-film stacks to investigate their interfacial structure and local strain state. Cross-sectional TEM lamellae were prepared by focused ion beam (FIB) milling using a Helios NanoLab 600i dual-beam system. The lamellae were subsequently analyzed using a cold field-emission gun (cold-FEG) JEOL JEM-F200 microscope equipped with a Gatan Ametek EELS Continuum spectrometer and post-filter K3 and Stela direct electron detectors.
High-angle annular dark-field scanning transmission electron microscopy (HAADF-STEM) imaging was employed to examine the microstructure of the multilayer stacks and to determine the thickness and continuity of the individual layers. Chemical characterization of the interfaces was performed by electron energy-loss spectroscopy (EELS). High-resolution TEM (HRTEM) imaging was used to investigate the atomic-scale structure of the interfaces using the K3 direct electron detector. Fast Fourier transform (FFT) and inverse FFT (IFFT) analyses were performed to identify localized deviations from the reference Si lattice periodicity and to visualize structurally perturbed regions in the vicinity of the interfaces (see Supplementary Information: HRTEM FFT/IFFT analysis).


\section*{DATA AVAILABILITY}

The data that support the findings of this study are available from the corresponding author upon reasonable request.


\section*{ACKNOWLEDGMENT}

We acknowledge funding through the Munich Quantum Valley project by the Free State of Bavaria, and the Munich Quantum Valley Quantum Computer Demonstrators - Superconducting Qubits (MUNIQC-SC) project (grant no. 13N16188), funded by the Federal Ministry of Research, Technology, and Space, Germany. We would also like to thank our colleagues from Fraunhofer EMFT for initial resonator measurements and shared  measurement code, TUM's Walter-Schottky Institute (WSI) and Center for Nanotechnology and Nanomaterials (ZNN) for shared use of their clean room facilities, the colleagues from Walther-Meissner Institute (WMI) for initial resonator layout and copper box design, the colleagues from TUM Catalysis Research Center (CRC) for giving us access to their XRD-measurement equipment, the TUM electron microscopy core facility (TUM\textit{em)} by providing access to the instrumentation for TEM, EDX and EELS analysis and the staff at TUM ZEITlab for STEM measurements and taking care of the lab facilities.  

\section*{AUTHOR CONTRIBUTIONS}

M.S., M.T. and B.S. conceived the idea. M.S. designed and carried out the experiments. M.T. provided the experimental facilities and managed the project. E.W. and A.O. carried out the TEM, EELS and CBED-STEM analysis. M.S., H.G. and M.T. analyzed the data and interpreted the study. M.S. wrote the manuscript. M.T., H.G., B.S., and E.W. revised the manuscript. All authors discussed the results and commented on the manuscript. 

\section*{COMPETING INTERESTS}

The authors declare no competing interests.


\printbibliography

\end{document}


\maketitle

\section{Atomic Force Microscopy}

For surface characterization, an area of 5~$\mu$m~x~5~$\mu$m of the thin films surfaces was scanned in tapping mode. A rate of 1~Hz and 512 lines per scan was used. The resulting height images are shown below in Fig.~\ref{AFM}. The chips were measured within the same day of deposition to avoid measurements on a fully grown native oxide.   

\begin{figure}[ht]
\begin{center}
  \includegraphics[width=0.9\textwidth]{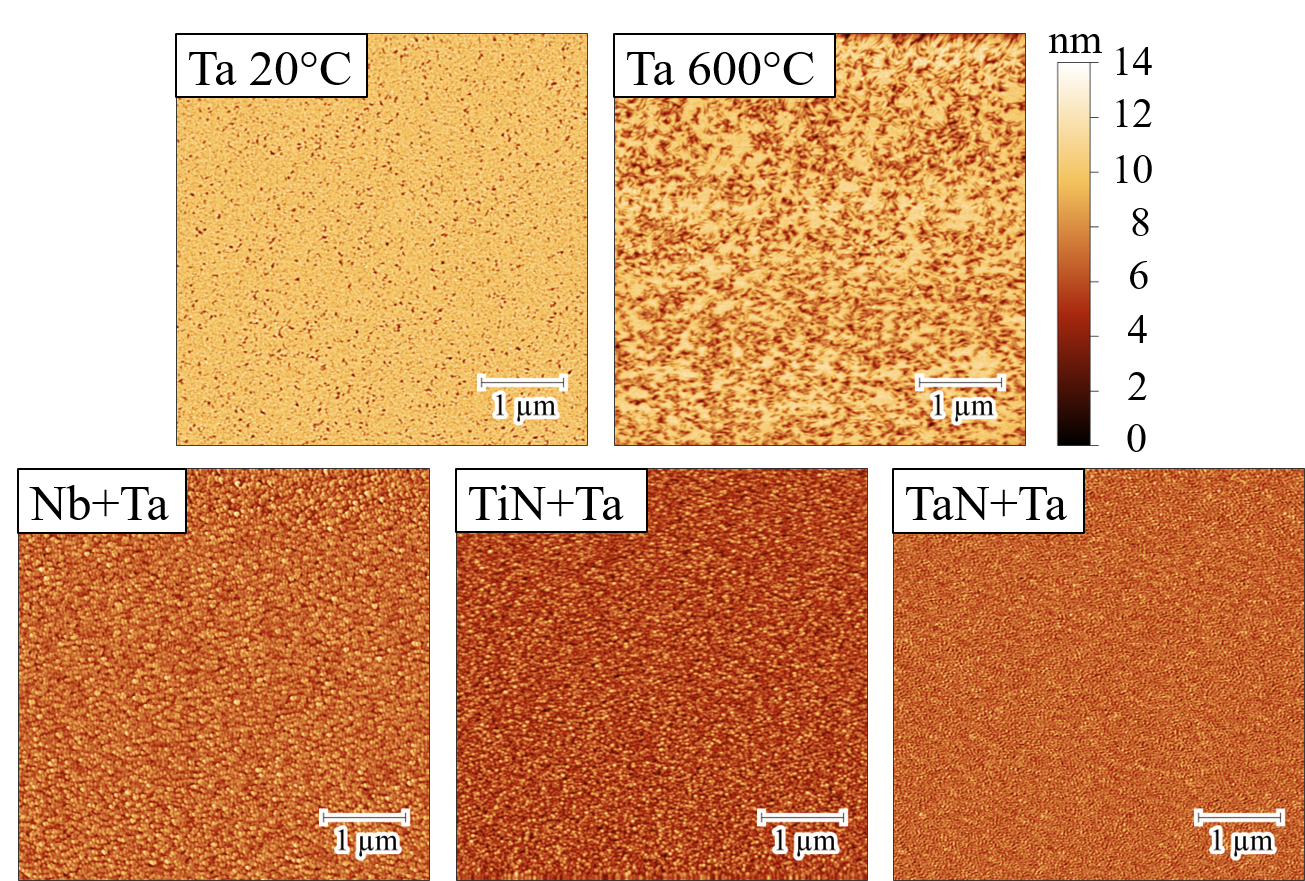}
  \caption{AFM pictures of the surfaces of the investigated thin films. Tantalum deposited at 20°C shows a smooth surface with an RMS roughness value (R\textsubscript{q}~=~0.5~nm) while the Ta sputtered at 600°C shows enlongated structures and a higher roughness (R\textsubscript{q}~=~1.5~nm). The Ta thin films sputtered on the seed layer show similar surface morphologies and roughness values Nb+Ta (R\textsubscript{q}~=~0.7~nm), TiN+Ta (R\textsubscript{q}~=~0.8~nm) and TaN+Ta (R\textsubscript{q}~=~0.8~nm).}
  \label{AFM}
  \end{center}
\end{figure}



\section{Influence of sputtering temperature on the growth of tantalum phases}

\begin{figure}[h!t]
\begin{center}
  \includegraphics[width=0.75\textwidth]{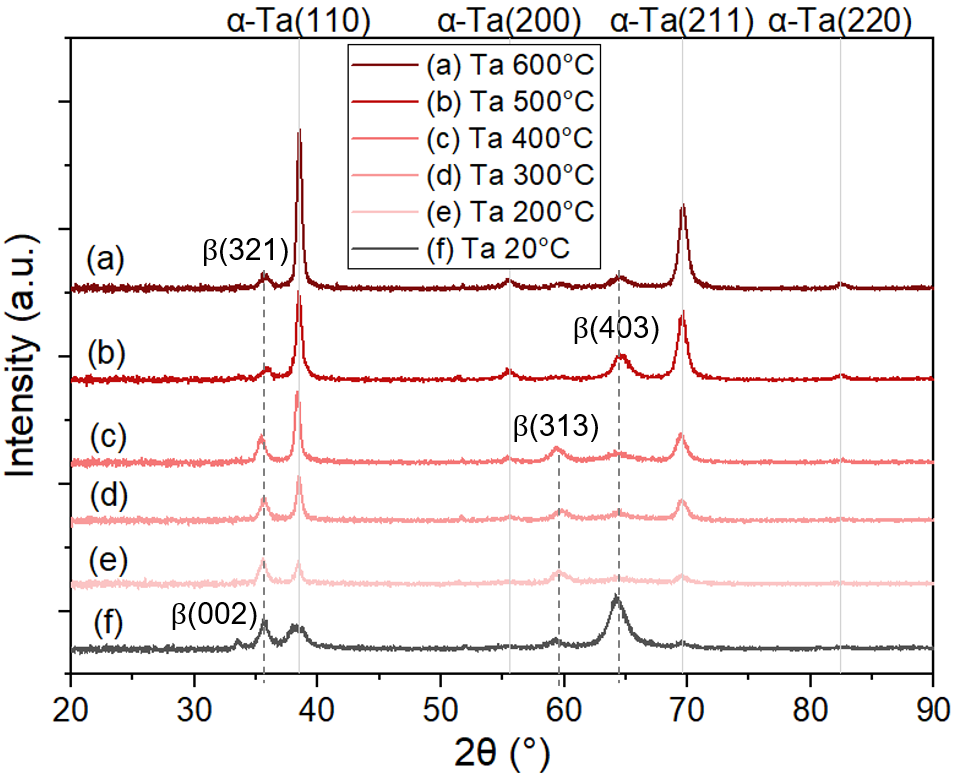}
  \caption{GI-XRD spectra of 200 nm Ta films deposited on Si(100) at substrate temperatures ranging from 600°C (a) to 20°C (f). With decreasing deposition temperature, the spectra show a gradual transition from predominantly $\alpha$-Ta to $\beta$-Ta, evidenced by the emergence and increasing intensity of characteristic $\beta$-Ta peaks ((002), (313), (321), (403)) and the concurrent reduction of $\alpha$-Ta reflections. The room-temperature deposited film (20°C) exhibits a $\beta$-Ta-dominated mixed phase, while high-temperature deposition favors the formation of $\alpha$-Ta.}
  \label{Ta_T}
    \end{center}
\end{figure}

In order to investigate the influence of the sputtering temperature on the growth of Ta phases, the same Si (100) substrates as already mentioned were heated to different temperatures and 200~nm thick Ta films were deposited. The Si chips were cleaned with the same routine, as outlined earlier. The obtained thin films were measured with GI-XRD and the results are shown in Fig.~\ref{Ta_T}. The top curve (a) represents the Ta sputtered on the 600°C heated substrate. As already seen before, in these films mostly alpha-Ta peaks are present. There are also two smaller peaks that cannot be attributed to alpha-Ta but to beta-Ta (321) and (403). At 500°C (b) the spectrum shows a more prominent peak for the beta-Ta (403), while the alpha peaks are still prominent and the beta (321) peaks remains. At 400°C the alpha (211) is noticeably lower and another beta-Ta peak (313) has emerged, while the beta (321) is more pronounced. This trend continues when looking at the 300°C (d) spectrum, where the alpha-Ta peaks are still visible, but beta-Ta peaks are as equally present. In the 200°C spectrum the alpha peaks are further diminished, especially the alpha (211) peak. 

As discussed before, the diffraction spectrum for the Ta sputtered at 20°C on Si without any intentional heating, exhibits only a weak contribution from the alpha‑Ta (110). Instead, the spectrum is dominated by features characteristic of the beta‑Ta phase, with identifiable peaks corresponding to the (002), (321), and (313) planes and a particularly intense peak associated with the beta‑Ta (403) orientation. 


\section{DC measurements of thin films}

\begin{figure}[ht]
\begin{center}
  \includegraphics[width=0.65\textwidth]{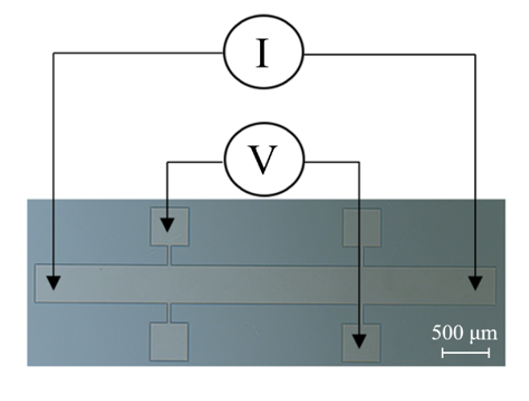}
  \caption{Optical image of the patterned Hall-bar structure used for resistivity and critical temperature measurements of the thin films. A constant current is applied along the 3600~$\mu$m long and 300~$\mu$m wide bar, while the voltage drop is measured between probes separated by 1500~$\mu$m, enabling accurate four-point resistance measurements.}
  \label{Hall}
  \end{center}
\end{figure}

To measure the resistivity of the sputtered thin films, a Hall-bar structure was used. The structure shown in Fig.~\ref{Hall} comprises a bar that is 3600~$\mu$m long and 300~$\mu$m wide. At a distance of 1500~$\mu$m apart there are pads to either side of the bar. For the measurements, a current of 100~$\mu$A is forced through the ends of the bar and the voltage drop is measured between two pads. As the thickness of the thin films is also known, the resistivity can be calculated from the measured resistance using Ohm's law. The structure was also used to measure the critical temperature of the thin films.   

\section{Additional resonator data}

\begin{figure}[ht]
\begin{center}
  \includegraphics[width=0.95\textwidth]{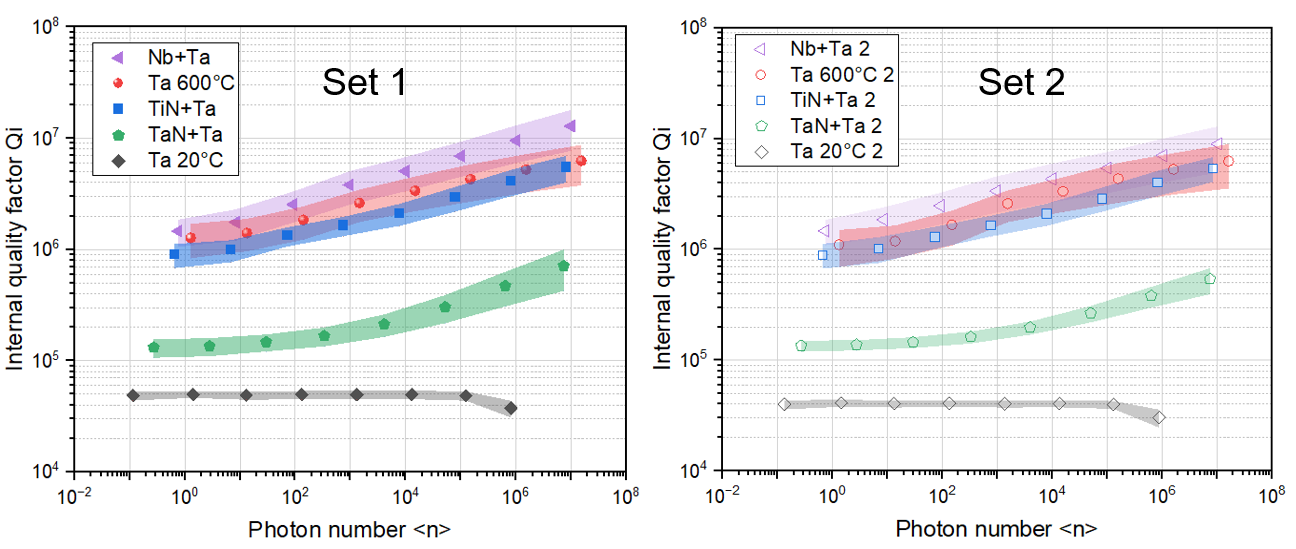}
  \caption{Internal quality factors extracted from the before-mentioned circle fits over photon number. Two sets of resonator chips were measured. After the initial set of this study, the five Ta thin films were fabricated again and measured as well (Set 2).}
  \label{ResDat}
  \end{center}
\end{figure}

The investigated thin films were measured in an initial first set of samples. After that, the cleaning, deposition and measurement procedure was kept the same and an additional set of samples were fabricated and measured. The results are shown in Fig. \ref{ResDat}, all measurements were carried out at 100~mK. Within the error margins, that are mainly due to the fitting procedure, the data is consistent in between the two runs. The trend and behavior of the thin films is also coherent with our previous study\cite{SINGER2024TantalumThinFilmsSputteredSiliconDifferentSeedLayersMaterialCharacterizationCoplanarWaveguideResonatorPerformance}.

The second set of measurements confirms the reproducibility of the observed trends in internal quality factors across all investigated thin-film stacks. The relative performance of the materials remains consistent, with Ta on Nb yielding the highest $Q_i$, while Ta on TaN consistently exhibits significantly lower values. Importantly, the agreement between independent fabrication runs indicates that the resonator performances are robust and not dominated by statistical fluctuations or processing variations. This strongly supports the conclusion that the differences in $Q_i$ originate from intrinsic material properties and, in particular, from the interfacial conditions discussed in the main text.

\section{Optimization of reactively sputtered nitrides}

As previously mentioned, tantalum nitride (TaN) and titanium nitride (TiN) were used, besides Niobium (Nb), as seed layers between the silicon substrate and the tantalum thin films. To achieve alpha-Ta growth on the reactively sputtered TaN and TiN layer, the process parameters were optimized to get cubic lattice TaN and TiN films of thickness 10 nm. 

In the case of TaN, sputtering was carried out at the base pressure of the chamber of 1$\cdot10^{-7}$~mbar, 200~W power, an argon flow rate of 20 sccm and a gas pressure of 18~$\mu$bar. The nitrogen flow rate was varied from 1~sccm to 12~sccm.

\begin{figure}[ht]
\begin{center}
  \includegraphics[width=1\textwidth]{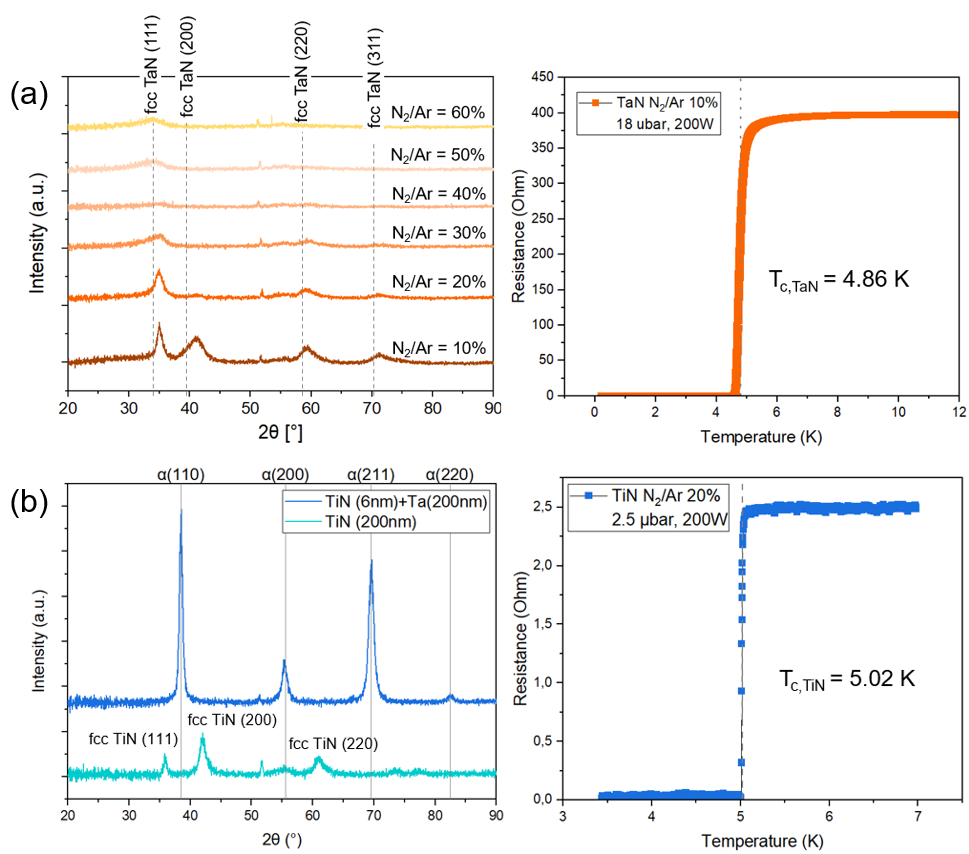}
  \caption{(a) GI-XRD spectra for TaN thin films sputtered with different nitrogen to argon contents and critical temperature measurement of the TaN thin film sputtered with a N\textsubscript{2}/Ar~=~10~\%. (b) GI-XRD spectra for TiN and Ta on TiN seed layer, showing alpha-Ta growth on cubic lattice TiN, and critical current measurement of the TiN thin film.} 
  \label{Nitrides}
  \end{center}
\end{figure}

The resulting GI-XRD spectra of the TaN thin films are shown in Fig. \ref{Nitrides}(a). The spectrum for a flow rate of a N\textsubscript{2}/Ar~=~60~\% only shows a peak around 35° of $\theta$ which can be attributed to fcc TaN (111). Besides that, only the sharp peak around 51-52° from the Si background is visible. The spectrum for flow rates of a N\textsubscript{2}/Ar~=~50~\% looks similar, while for the TaN with a N\textsubscript{2}/Ar~=~40~\% the (111) peak is diminished. When the nitrogen flow rate is further decreased, peaks corresponding to TaN (220) and (311) appear in the spectrum for N\textsubscript{2}/Ar~=~20~\%. When further decreasing the nitrogen to N\textsubscript{2}/Ar~=~10~\%, these peaks become more pronounced and the (111) peak becomes sharp and a peak (200), can also be observed. The nitrogen to argon flow ratio of~10~\% was subsequently used for sputtering the TaN thin films. For this thin film, a critical temperature measurement was carried out. The resistance over temperature is shown in Fig. \ref{Nitrides} on the right. It is observed that the film does turn superconducting. The transition starts slowly above 5~K and then drops sharply until the film is superconducting at ~4.7K. Because these measurements were done on Hall bar structures, the resistivity of the TaN film was extracted directly from the transport data. The resulting resistivity is 702~$\mu \Omega \cdot \textrm{cm}$.

For TiN, it was already known from our previous studies \cite{schoof_development_2024-1}, that sputtering at low pressure is required to get stoichiometric films. Therefore, sputtering was done at 2.5~$\mu$bar and with a flow rate of N\textsubscript{2}/Ar~=~20~\%, the rest of the parameters were identical as described before. The GI-XRD spectrum of a 200~nm thick TiN film is shown and compared to 200~nm of Ta on a 6~nm thin TiN film in Fig. \ref{Nitrides}(b). In the spectrum for TiN we can identify fcc-TiN peaks (111), (200) and (220). The spectrum for the Ta on TiN shows only peaks at position of alpha-Ta, indicated by the vertical reference lines, confirming that we grow a cubic TiN that nucleates the growth of alpha-phase Ta. The critical temperature of the 200~nm TiN film is 5.02~K.

\section{Lattice constant and strain calculation from XRD-data}

\begin{figure}[ht]
\begin{center}
  \includegraphics[width=1.05\textwidth]{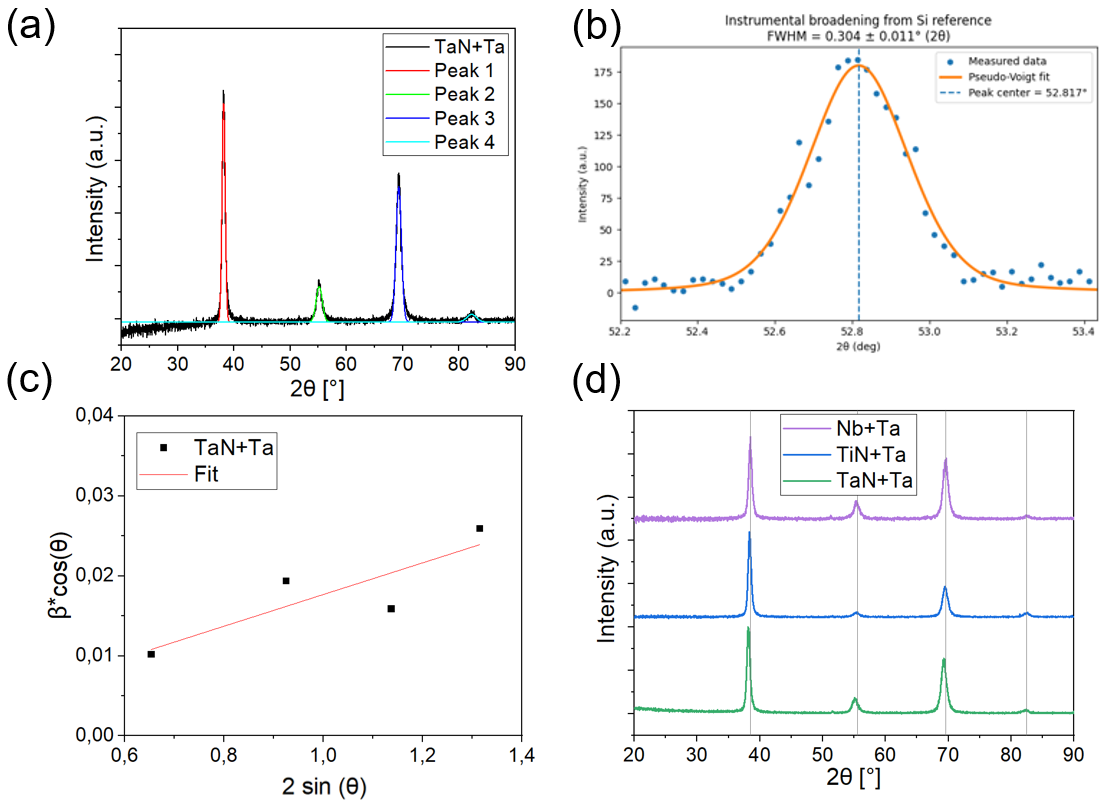}
  \caption{(a) Measured XRD data from a Nb+Ta thin film and fitted Gaussian curves. (b) Fit from a Si reference measurement to determine the line broadening of the instrument.(c) Exemplary linear Williamson-Hall fit of a Ta on TaN seed layer thin film. (d) Comparison of XRD spectra of 10~nm thick Ta sputtered on Nb, TiN and TaN seed layers. The peaks of Ta on TaN are shifted the most from their reference positions, corresponding to the crystal planes (110), (200), (211) and (220), from left to right.}
  \label{XRD_Fit}
  \end{center}
\end{figure}

In order to extract peak positions $\theta$ and FWHM $\beta$ from the GI-XRD measurements, a Gaussian fit is applied to the data, as shown exemplary for data from a Nb+Ta thin film in Fig. \ref{XRD_Fit}(a). A correction for line broadening of the instrument $\beta_{\mathrm{inst}}$ is carried out by using the formula \cite{cullity_elements_1978,mote_williamson-hall_2012}

\begin{equation}
\beta_{\mathrm{meas}}^{2}
=
\beta_{\mathrm{sample}}^{2}
+
\beta_{\mathrm{inst}}^{2}
\end{equation}

to calculate the actual broadening from the sample $\beta_{\mathrm{sample}}$ from the measured value $\beta_{\mathrm{meas}}$. The line broadening of the instrument was determined with a reference Si sample to be $0.304^\circ \pm 0.011^\circ$ with the used grazing incidence set-up of $0.5^\circ$, as shown in \ref{XRD_Fit}(b). The used Si peak that appears at around $52.8^\circ$ in the spectrum is usually forbidden, but is visible here because of the configuration of the grazing incidence angle. As mentioned in the main text, the Williamson-Hall formula 

\begin{equation}
\beta cos \theta=\frac{K_L\lambda}{D} + K_D 2\varepsilon sin \theta
\end{equation}

 is used to fit the data points with a linear curve as shown exemplarily in Fig. \ref{XRD_Fit}(c). We do not assume a certain value for the constant factor $K_D$, so the extracted $\varepsilon$ multiplied by $K_D$ is determined by the slope of the fit \cite{JIANG1999applicabilityxraydiffractionlineprofileanalysisextractinggrainsizemicrostrainnanocrystallinematerials, MOHAMADZAIDI2019CrystallitesizemicrostrainXRDlinebroadeninganalysisAgSiNthinfilms, KHORSANDZAK2011XrayanalysisZnOnanoparticlesWilliamsonHallsizestrainplotmethods, MAEvaluationlatticestrainZnOthinfilmsbasedWilliamsonHallanalysis, BIRKHOLZThinFilmAnalysisXRayScattering}. 

The term $\frac{K_L \lambda}{D}$ is determined by the intercept of the linear fit with the y-axis. Assuming $K_L$ to be 0.9 and with an x-ray wavelength of 0.154~nm, the crystallite size $D$ was calculated. The Ta on Nb seed layer has the smallest crystallite size $D_{Ta+Nb}$~=~12.7~$\pm$~5.2~nm. The Ta on TiN seed layer has a slightly larger crystallite size $D_{Ta+TiN}$~=~19.4~$\pm$~9.5~nm while the Ta on the TaN seed layer shows the largest one $D_{Ta+TaN}$~=~64.2~$\pm$~18.4~nm. \\

To verify that the assigned crystal planes to the peaks in the XRD-data are correct, the lattice constants can be calculated and compared to literature values. For this purpose, Bragg's law can be used which states that 

\begin{equation}
n\cdot\lambda=2d\cdot sin(\theta) 
\end{equation}

where $n$ is the diffraction order. By re-arranging the formula, the grating constant $d$ can be calculated. Using the formulas for tetragonal 

\begin{equation}
\frac{1}{d^2}=\frac{h^2+k^2}{a^2}+\frac{l^2}{c^2}
\end{equation}

and cubic

\begin{equation}
\frac{1}{d^2}=\frac{h^2+k^2+l^2}{a^2}
\end{equation}

crystal structures \cite{BIRKHOLZThinFilmAnalysisXRayScattering} with the Miller indices $h$, $k$ and $l$ the lattice constants $a$ and $c$ can be calculated by choosing a crystal plane where either $h$ and $k$ or $l$ is zero, eliminating one term and allowing for the other one to be determined. With this procedure we find the following lattice constants shown in Tab. \ref{Lat}, which are in agreement with values from literature. Discrepancies between calculated and literature values may be due to peak shifts and subsequent error progression while calculating $d$. This effect may not be visible in some cases due to the anisotropic nature of the strain in thin films, resulting in different peak shifts. \\

The macroscopic strain in the thin films results in shifts of peaks in the XRD spectrum. As shown in the main text, the peak shifts correlate to the microstrain determined by the Williamson-Hall method. The full spectra of the thin 10~nm thick Ta layer on the seed layers Nb, TiN and TaN are shown in Fig.~\ref{XRD_Fit}(d). The four vertical lines represent references for alpha-Ta peaks of unstrained Ta.

\begin{table}[ht!]
\centering
\caption{Calculated lattice constants from XRD-data and literature values \cite{HALLMANN2013Effectsputteringparameterssubstratecompositionstructuretantalumthinfilms, HULPKE1992Latticedynamicsniobium001surface, IGASAKI1978StructureElectricalPropertiesTitaniumNitrideFilms}.}
\begin{tabular}[t]{lccc}
\hline
Material&Crystal structure&Calculated lattice&Lattice constant \\
&&constant [Å]&in literature [Å]\\
\hline
$\alpha$-Ta (Nb+Ta)& body-centered cubic & 3.307 & 3.306 \\
&&&\\
$\beta$-Ta (Ta 20°C)& tetragonal & a=5.326 & a=5.313\\
&  & c=10.814 & c=10.194\\
&&&\\
TiN& face-centered cubic & 4.329 & 4.241 \\
&&&\\
TaN& face-centered cubic & 4.334 & 4.336\\
&&&\\
Nb&body-centered cubic &3.324&3.300\\
\hline
\end{tabular}
\label{Lat}
\end{table}%

\section{Determination of lattice distortions from HRTEM pictures}



\begin{figure}[ht]
\begin{center}
  \includegraphics[width=1\textwidth]{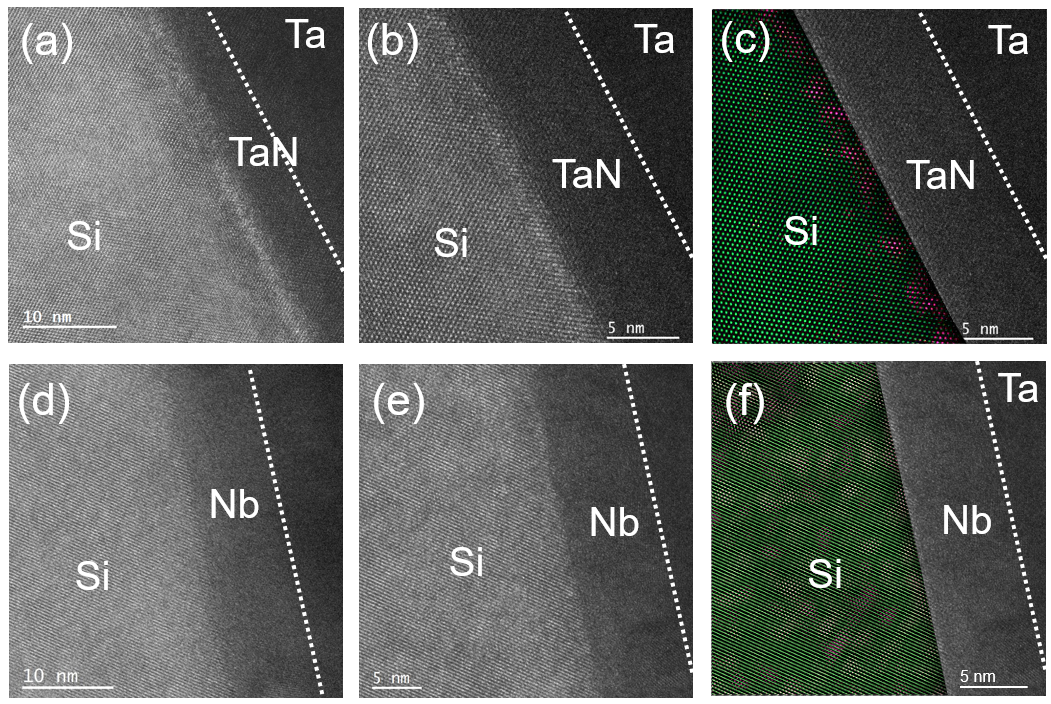}
  \caption{HRTEM pictures of Ta on TaN slightly zoomed out (a) and further zoomed in (b) towards the metal-to-substrate interface. FFT analysis of the zoomed in picture (c) with overlaid Si lattice up to the interface to the seed layer. Green areas denote an unstrained  Si lattice, while pink areas mark distorted lattice regions.  Analogue, HRTEM pictures of Ta on Nb (d)+(e) and corresponding FFT pattern of the Si lattice close to the interface (f).}
  \label{FFT}
  \end{center}
\end{figure}

Figure S7 compares the atomic-scale structure of the TaN/Si and Nb/Si interfaces using HRTEM imaging and FFT/IFFT analysis. The HRTEM micrographs of the TaN/Si interface (Fig.~\ref{FFT}(a),(b)) exhibit localized contrast variations extending along the interface, whereas the Nb/Si interface (Fig.~\ref{FFT}(d),(e)) appears comparatively uniform. Since HRTEM contrast originates from a combination of diffraction and phase-contrast effects, these variations in contrast cannot be directly interpreted as lattice distortions. Therefore, FFT/IFFT analysis was performed to identify regions where the local lattice periodicity deviates from that of the reference Si crystal. It should be noted that this procedure does not provide a quantitative strain value, as obtained by geometric phase analysis, but rather serves as a qualitative method to spatially identify lattice perturbations relative to the defect-free Si reference. The localization of the residual IFFT signal at the interface therefore indicates the presence of an interfacial structurally disturbed region, which may arise from strain, lattice defects, or a combination thereof.

First, a fast Fourier transform (FFT) was calculated from the entire image region containing both the Si substrate and the TaN interface. To isolate the contribution of the undistorted Si lattice, an FFT was separately obtained from a defect-free Si region away from the interface, where the crystal periodicity was assumed to represent the unstrained reference structure. The reference Si FFT was then subtracted from the FFT of the entire image to suppress the contribution from the bulk Si periodicity and retain only the residual frequency components associated with deviations from the ideal Si lattice. These residual components were attributed to interfacial lattice distortions and strain-induced modifications of the local periodicity. Subsequently, inverse FFTs were performed using (i) the selected reflections corresponding to the unstrained Si lattice and (ii) the residual FFT obtained after subtraction. The IFFT pattern reconstructed from the residual frequency components were overlaid onto the Si side of the original HRTEM image and revealed that the structural perturbation was spatially localized at the TaN/Si boundary, indicating that the deviation from the ideal Si lattice is confined to the interfacial region (Fig.~\ref{FFT}(c)). In contrast, the Nb/Si interface (Fig.~\ref{FFT}(f)) exhibits substantially weaker localization of residual frequency components. These observations suggest the presence of a structurally perturbed region at the TaN/Si interface, consistent with localized strain and/or lattice distortion within the Si substrate.

\section{Cross sectional HAADF-STEM and EELS analysis of thin films}

\begin{figure}[ht]
\begin{center}
  \includegraphics[width=1\textwidth]{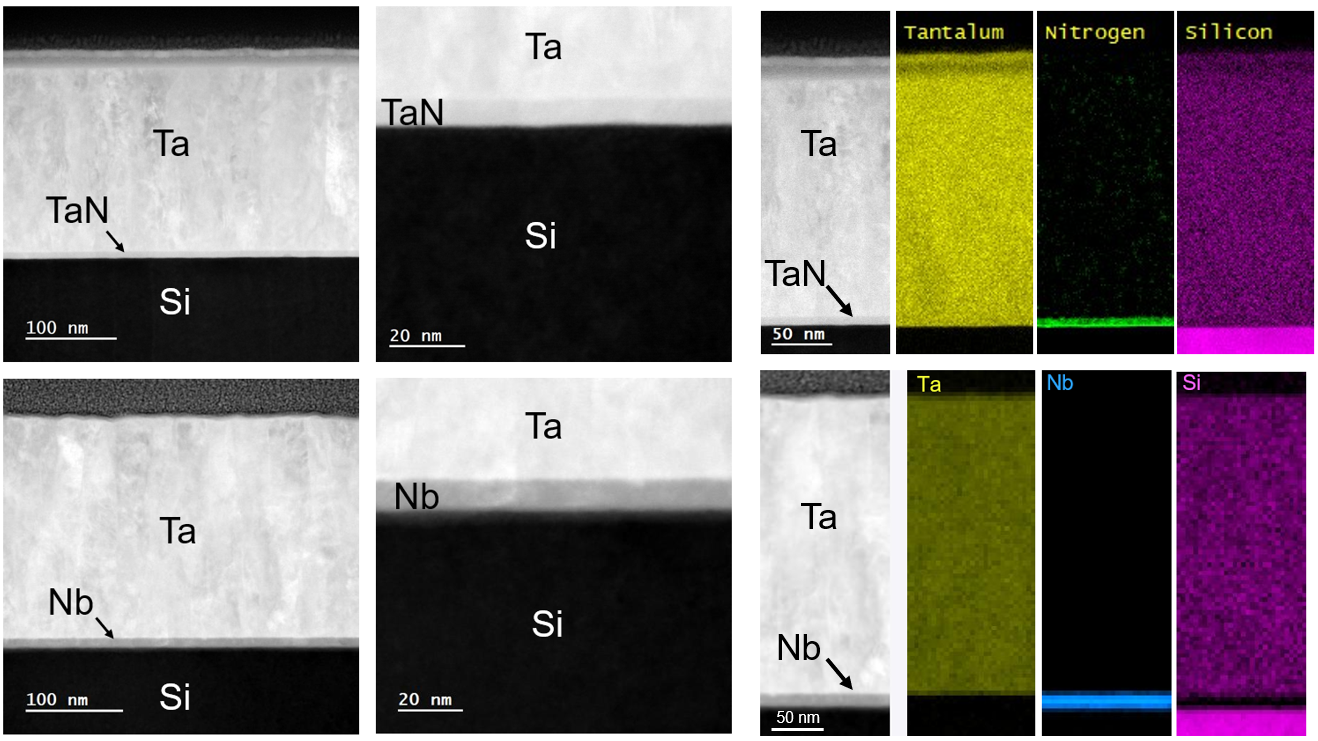}
  \caption{HAADF-STEM pictures and EELS analysis of cross sectional Ta on TaN (top row) and Ta on Nb (bottom row). The seed layer looks well defined and uniform; from the EELS pictures no inter-diffusion can be discerned.}
  \label{EELS}
  \end{center}
\end{figure}

FIB-TEM lamellas were prepared using focused ion beam (FIB) milling (Helios NanoLab 600i). Structural and chemical analyses were performed using a JEOL F200 scanning transmission electron microscope equipped with the Gatan EELS continuum spectrometer and the post-filter Stela direct electron detector. Representative HAADF-STEM images and corresponding EELS elemental maps are shown in Fig.~\ref{EELS}, where the top row corresponds to Ta deposited on a TaN seed layer and the bottom row to Ta deposited on an Nb seed layer.

The HAADF-STEM images reveal the approximately 200 nm thick Ta layer deposited on the Si substrate. A continuous interfacial layer is observed between Ta and Si, corresponding to the TaN or Nb seed layer. The measured contrast confirms the presence of the seed layers and indicates a thickness in the nanometer range consistent with the deposition design. The rough surface visible above the Ta layer originates from the protective layers deposited during TEM sample preparation. Prior to FIB milling, a protective carbon layer was deposited and the lamella was subsequently thinned using a Ga-ion beam.

The EELS elemental maps further confirm the chemical identity of the seed layers. For the TaN sample, nitrogen is localized at the TaN interlayer, whereas for the Nb-seeded sample, Nb is confined to the corresponding interfacial layer. In both cases, the seed layers appear laterally continuous and well defined. No significant elemental intermixing across the interfaces is observed within the spatial resolution and detection limits of the measurements, including from higher-magnification STEM-EELS line profiles across the interfaces (data not shown). The Si signal exhibits a weak apparent extension into the Ta-containing region, because of a partial overlap of Ta and Si signals and not due to a physical presence of Si inside the Ta film.

\section{Coplanar waveguide resonator design}

\begin{figure}[ht]
\begin{center}
  \includegraphics[width=1\textwidth]{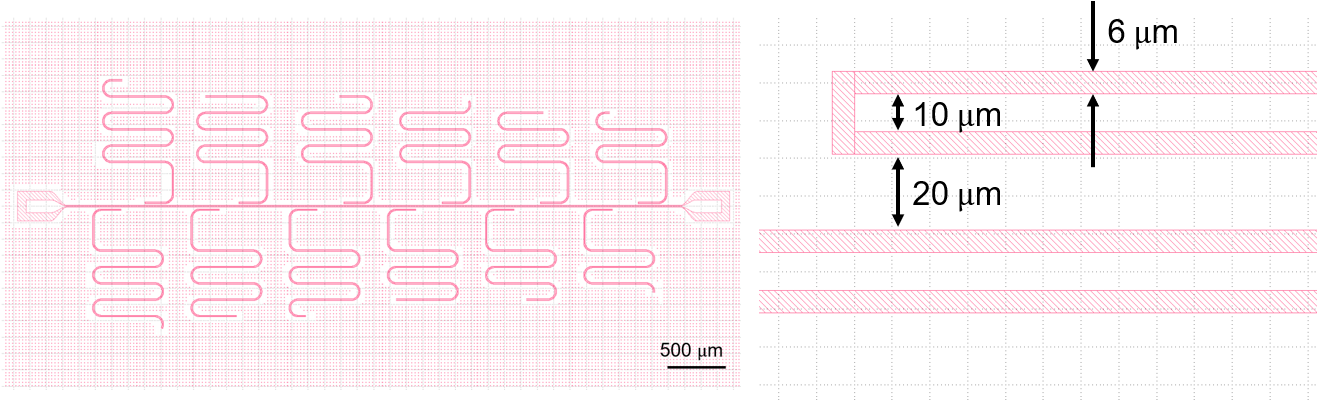}
  \caption{Design of the coplanar waveguide resonators. There are twelve resonators with different lengths on alternating sides of a feed line. The center line strip has a width of 10~$\mu$m, the gap width is 6~$\mu$m and the resonators are 20~$\mu$m spaced from the feed line.}
  \label{ResDesign}
  \end{center}
\end{figure}

The chips were patterned into coplanar waveguide resonators with varying lengths to target different resonance frequencies. To achieve this, a design with twelve resonators was created. As can be seen in Fig. \ref{ResDesign}, a common feed line runs through the chip. On either end of the feed line there is a pad to which bond wires can be connected and an RF-signal can be coupled into the chip. On alternating sides of the feed line, the resonators are located. One end of each resonator is open while the other one is shorted to create $\lambda$/4 resonators. 

The feed line and the resonators have a center line width of 10~$\mu$m and a gap width of 6~$\mu$m. The coupling part of the resonator, i.e., the part in close proximity to the feed line, is 200~$\mu$m long and there is a 20~$\mu$m distance between the coupling part and the feed line. With these design parameters the resonators have a resonant frequency in the range of 5-9~GHz and the coupling quality factor is around 2$\cdot10^5$.

\section{Cryogenic measurement set-up}

\begin{figure}[ht]
\begin{center}
  \includegraphics[width=1\textwidth]{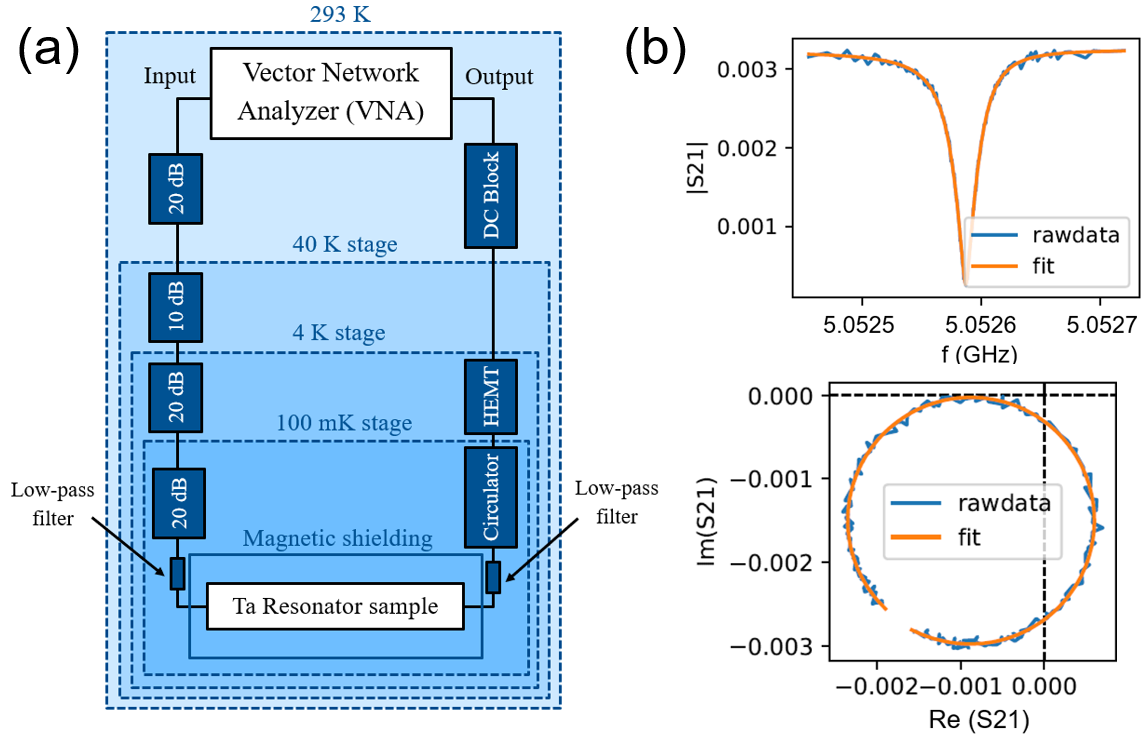}
  \caption{(a) Set-up of the cryogenic RF-measurements with a vector network analyzer. (b) Measured frequency dip in the S21-spectrum at low power of a coplanar-waveguide resonator at the top and corresponding circle in the complex plane at the bottom, both with fitted curves.}
  \label{Cryo_fit}
  \end{center}
\end{figure}

For cryogenic RF-measurements, an adiabatic demagnetization cryostat (kiutra, L-Type Rapid) was used. A vector network analyzer (Keysight, P5002B) was used to send RF-signals into the cryostat, with the sample chip mounted in the copper box typically held at the cryostat base temperature of 100 mK . All room temperature attenuators and cables amount to 40~dB attenuation, followed by additional attenuation of 10~dB at the 40~K stage, 20~dB at the 4~K stage and 20~dB at the sample stage of the cryostat. There, the attenuated signal passes through a low-pass microwave filter and is coupled into the copper box via SMA-connectors. Through a strip line on a PCB it finally reaches the sample via aluminum bond wires that are attached to the chip. The input signal goes into one end of the feed line of the chip, passes by all of the twelve CPW resonators, capacitively coupled to the feed line. The transmitted signal at the end of the feed line goes via another SMA-connector through a circulator and is subsequently amplified by a high electron mobility transistor (HEMT) and then fed back into the VNA as shown in Fig. \ref{Cryo_fit}(a).

For DC measurements, the same cryostat is used. There are 40 DC lines that go into the cryostat and onto the sample stage, 20 of which are equipped with an additional audio and radio frequency low-pass filter from QDevil. Signals are sent into the cryostat by  Keithley SMU 2635A/B devices.

\section{Quality factor measurements of resonators}

To find the resonance frequencies of the resonators, a coarse frequency sweep is carried out ranging from 4~GHz to 9~GHz in steps of 1~kHz. After this spectrum analysis, a finer frequency sweep is done around the found frequencies with a step size of 200~Hz and a power of -30~dBm, which is then followed by the actual power sweep measurement. The power is swept from -90~dBm to -160~dBm in -10~dB steps. At low powers, the measurement is carried out multiple times and the results are averaged as the data becomes noisier in that regime. For example, the measurement at -160~dBm is averaged over seven sweeps. 

An example for a resulting peak in the S21-spectrum of a Nb+Ta thin film is shown in Fig. \ref{Cryo_fit}(b) on top. The data is then mapped into the complex plane which creates a non-closed circle as shown in Fig. \ref{Cryo_fit}(b) on the bottom. The circle is not perfectly closed due to impedance mismatches, cable delays, and residual background transmission in the measurement setup, which introduce asymmetry and rotation in the complex response. The shown measurement is taken at low input power, where the signal-to-noise ratio is reduced due to the limited number of photons in the resonator. Consequently, the data appears noisier, but the underlying resonance response can still be reliably fitted, as indicated by the agreement with the fitted curve (orange). To improve statistical accuracy in this regime, multiple measurements are performed and averaged. With the detected resonance frequency, the loaded and coupling quality factors are determined and the internal quality factor is calculated as discussed in the main text.

\printbibliography